\documentclass[prb,longbibliography]{revtex4-1}
\pdfoutput=1
\usepackage{amsmath,amssymb,latexsym,epsfig,graphics,epsf}
\usepackage{graphics}
\usepackage{float}

\newcommand{\bR}{\bold R}

\newcommand{\bRa}{{\bf R}_{\rm a}}

\newcommand{\bRb}{{\bf R}_{\rm b}}

\newcommand{\tbR}{\tilde{\bold R}}

\newcommand{\be}{\begin{equation}}
\newcommand{\ee}{\end{equation}}
\newcommand{\fig}[1]{Fig.~\ref{#1}}
\newcommand{\eq}[1]{Eq.~(\ref{#1})}

\newcommand{\Sex}{{S}_{\rm ex}}

\graphicspath{ {FinalFigures/} }
\usepackage{lpic}

\begin{document}

\title{Density-scaling exponents and virial potential-energy correlation coefficients for the $(2n,n)$ Lennard-Jones system}\date{\today}
\author{Ida M. Friisberg}\email{idafriisberg@gmail.com}
\author{Lorenzo Costigliola}\email{lorenzo.costigliola@gmail.com}
\author{Jeppe C. Dyre}\email{dyre@ruc.dk}
\affiliation{``Glass and Time'', IMFUFA, Department of Science and Environment, Roskilde University, P.O. Box 260, DK-4000 Roskilde, Denmark}

\begin{abstract}
This paper investigates the relation between the density-scaling exponent $\gamma$ and the virial potential-energy correlation coefficient $R$ at several thermodynamic state points in three dimensions for the generalized $(2n,n)$ Lennard-Jones (LJ) system for $n=4, 9, 12, 18$, as well as for the standard  $n=6$ LJ system in two, three, and four dimensions. The state points studied include many low-density states at which the virial potential-energy correlations are not strong. For these state points we find the roughly linear relation $\gamma\cong 3nR/d$ in $d$ dimensions. This result is discussed in light of the approximate ``extended inverse power law'' description of generalized LJ potentials [N. P. Bailey \textit{et al.}, J. Chem. Phys. \textbf{129}, 184508 (2008)]. In the plot of $\gamma$ versus $R$ there is in all cases a transition around $R\approx 0.9$, above which $\gamma$ starts to decrease as $R$ approaches unity. This is consistent with the fact that $\gamma\rightarrow 2n/d$ for $R\rightarrow 1$, a limit that is approached at high densities and/or high temperatures at which the repulsive $r^{-2n}$ term dominates the physics.
\end{abstract}

\maketitle

\section{Introduction}\label{seq:introduction}

In the past decade a class of systems has been identified that is well described by the isomorph theory \cite{paper1,paper2,paper3,paper4,paper5,simpleliquid,hiddenscale,JeppeBigReview}, sometimes referred to as R (Roskilde) simple systems \cite{mal13,abr14,fer14,fle14,pra14,Isomorph2.0,buc15,har15,hey15,sch15}. This class is believed to include most van der Waals bonded and metallic liquids and solids, as well as most weakly dipolar or ionic systems \cite{paper1,paper2,paper3,paper4,paper5,simpleliquid,hiddenscale,JeppeBigReview}. The class does not include systems with strong directional bonds like hydrogen or covalently bonded systems \cite{JeppeBigReview}. 

To determine whether the isomorph theory holds for a given system one calculates the strength of the virial potential-energy correlations at the  state points in question in $NVT$ simulations, i.e., in the canonical ensemble \cite{paper1,paper2}. Isomorphs are curves of invariant  structure and dynamics in the thermodynamic phase diagram, see Sec. \ref{sec:isomorph}. A system has isomorphs if and only if it has strong virial potential-energy correlations \cite{paper4}. Virial potential-energy correlations are quantified by the state-point dependent Pearson correlation coefficient $R$ defined as

\begin{align}\label{eq:correlation}
R\left(\rho, T\right) = 
\frac{\langle\Delta W \Delta U \rangle}{\sqrt{\langle\left(\Delta W\right)^{2} \rangle\langle\left(\Delta U\right)^{2}}\rangle}\,.
\end{align}
In \eq{eq:correlation} $\rho$ is the (number) density ($\rho=N/V$ with $N$ particles in volume $V$), $T$ is the temperature, $W$ and $U$ are the virial and potential energy, the brackets denote $NVT$ averages, and $\Delta$ is the instantaneous deviation from equilibrium mean values.

The isomorph theory is exact only for systems with a potential that is a constant plus an Euler-homogeneous function, in which case the correlation coefficient $R$ in \eq{eq:correlation} is unity for all $\rho$ and $T$ \cite{paper4}. An example is the inverse-power-law (IPL) pair-potential systems. For non-Euler-homogeneous potentials, which includes all realistic systems, the quantity $R$ is less than unity. In that case the applicability of the isomorph theory is restricted to a certain region of the phase diagram, usually the condensed-phase region encompassing the solid and ``ordinary'' liquid phase. 

A system is defined to be R simple whenever $R>0.9$ \cite{paper1}, because this ensures that isomorphs exist to a good approximation \cite{paper4}. The $R=0.9$ threshold is somewhat arbitrary, however. The present paper investigates what happens when $R$ falls below this threshold. From the isomorph theory there is little help, because it generally breaks down when correlations are no longer strong, i.e., when the correlation coefficient goes significantly below $0.9$ \cite{paper2,paper4,hiddenscale}. In 2014, however, an interesting paper appeared by Prasad and Chakravarty \cite{pra14} establishing that Rosenfeld's excess entropy scaling as well as density-scaling \cite{rol05}, which may both be derived from the isomorph theory \cite{paper4}, need not break down even if the virial potential-energy correlations are weak. Reference \onlinecite{pra14} studied the transition to simple liquid behavior in computer simulations of modified water models. We have taken inspiration from this work to systematically study the region of the phase diagram where virial potential-energy correlations are not strong for a class of models that -- in contrast to water models -- do have sizable regions of strong correlations in the thermodynamic phase diagram. This is done by studying Lennard-Jones type systems at lower densities than have previously been in focus.

This paper presents a study of the LJ($2n,n$) class of potentials defined as single-component systems with LJ-type radially symmetric pair potentials with a repulsive term proportional to $r^{-2n}$ and an attractive term proportional to $r^{-n}$ ($r$ being the interparticle distance). Results are presented for the cases $n=4, 9, 12, 18$ in three dimensions, as well as for the standard LJ potential ($n=6$) in two, three, and four spatial dimensions (Sec. \ref{sec:results}). Before doing this the isomorph theory is briefly reviewed (Sec. \ref{sec:isomorph}). Section \ref{sec:disc} summarizes the paper's findings.

\section{Theoretical/Computational}\label{sec:isomorph}

\subsection{Aspects of the isomorph theory}

R simple systems -- previously termed ``strongly correlating'' which, however, led to confusion with strongly correlated quantum systems -- were defined in 2009 by reference to strong virial potential-energy correlations \cite{paper1}. This section presents the alternative definition of the same class of systems given in 2014 \cite{Isomorph2.0} that refers to their ``hidden scale invariance'' \cite{hiddenscale}. The original formulation of the isomorph theory \cite{paper4} is recovered via a first-order Taylor expansion \cite{Isomorph2.0}. 

Consider an $N$-particle system in $d$ spatial dimensions. The system is defined to be R simple if for any two configurations corresponding to the same density that obey $U(\bRa)<U(\bRb)$, this ordering is preserved after a uniform scaling of the two configurations \cite{Isomorph2.0} ($\bR$ is the $d\,N$ dimensional vector describing a configuration of $N$ particles in $d$ spatial dimensions). Formally, this condition is written \cite{Isomorph2.0}

\begin{align}\label{eq:scaling}
U\left(\bRa\right) < U\left(\bRb\right) 
\Rightarrow 
U\left(\lambda\bRa\right) < U\left(\lambda\bRb\right) \,.
\end{align}
In computer simulations periodic boundary conditions are applied, and it is understood that the box size scales with $\bR$. Equation (\ref{eq:scaling}) need not be obeyed for all configurations; as long as it applies for most of the physically relevant configurations, the predictions of isomorph theory are obeyed to a good approximation.
In \eq{eq:scaling} the scaling factor $\lambda$ can be any positive real number, and the equation holds for scaling "both ways". Because of this an R simple system obeys \cite{Isomorph2.0}

\begin{align}\label{eq:scaling2}
U\left(\bRa\right) = U\left(\bRb\right) 
\Rightarrow 
U\left(\lambda\bRa\right) = U\left(\lambda\bRb\right)\,.
\end{align}
Thus if two configurations have the same potential energy, their scaled potential energies are also identical. 

R simple systems are characterized by strong correlations between virial and potential-energy fluctuations in the $NVT$ ensemble. This can be derived from \eq{eq:scaling2} in the following way. For two configurations, $\bRa$ and $\bRb$, at the same density with the same potential energy, \eq{eq:scaling2} states that $U(\lambda \bRa)=U(\lambda \bRb)$. Taking the derivative of this identity with respect to $\lambda$ one obtains

\begin{equation}
\bRa \cdot \nabla U(\lambda\bRa) = \bRb \cdot \nabla U(\lambda\bRb)\,.
\end{equation}
Using the definition of virial $W(\bf{R}) \equiv {\bf{R}} \cdot \nabla U(\bf{R})/{\it d}$ \cite{han13}, one gets for $\lambda = 1$

\begin{equation}
W(\bRa) = W(\bRb)\,.
\end{equation}
Thus $U(\bR)$ determines $W(\bR)$, implying perfect correlation. This applies for systems that satisfy \eq{eq:scaling} or, equivalently \eq{eq:scaling2}, for all configurations; it holds for realistic R simple systems to a good approximation. As a consequence, for such systems the correlation coefficient $R$ of \eq{eq:correlation} is close to unity, but not exactly unity. As mentioned, the threshold defining R simple systems has usually been taken to be $R=0.9$ \cite{paper1}.

Two configurations of density $\rho_1$ and $\rho_2$, which scale uniformly into one another, i.e., $\bR_{2}=\lambda\bR_{1}$ for some $\lambda$, are related through the following equality (where $d$ is the dimension):

\begin{align}\label{eq:reducedCoordinate}
\rho_{1}^{1/d}\boldsymbol{R}_{1} = \rho_{2}^{1/d}\boldsymbol{R}_{2} \equiv \tilde{\boldsymbol{R}} \,.
\end{align}
The last equality defines the ``reduced'' (dimensionless) configuration vector $\tbR$. Henceforth reduced quantities are marked by a tilde. The reduced version of a physical quantity is obtained by using the ``macroscopic'' unit system in which the length unit is $\rho^{-1/d}$, the energy unit is $k_BT$, and the time unit is $\rho^{-1/d}/\sqrt{mk_BT}$ ($m$ is the particle mass) \cite{Rosenfeld1999}.

The entropy of any physical system can be written as an ideal-gas term plus an excess term $S_{\text{ex}}$. The ideal-gas term is the entropy of an ideal gas at the density and temperature in question, the excess term is the contribution to entropy from the interactions between the particles. We define the microscopic excess entropy function, $S_{\text{ex}}({\bf R})$, to be the thermodynamic excess entropy \cite{Isomorph2.0, LandauLif} of the potential-energy surface the configuration belongs to, i.e.,

\begin{align}\label{eq:excessEntropy}
S_{\text{ex}}(\boldsymbol{R}) \equiv S_{\text{ex}}\left(\rho, U\left(\boldsymbol{R}\right)\right)\,.
\end{align}
The function on the right-hand side, $S_{\text{ex}}\left(\rho, U\right)$, is the \textit{thermodynamic} excess entropy of the state point with density $\rho$ and average potential energy $U(\bR)$. In other words, the microscopic excess entropy of a configuration $\bR$ is defined as the excess entropy of a thermodynamic equilibrium system of same density with average energy precisely equal to $U(\boldsymbol{R})$.

Inverting \eq{eq:excessEntropy} we get

\begin{align}\label{eq:potEnergyExEntropy}
U\left(\boldsymbol{R}\right) = U\left(\rho, S_{\text{ex}}\left(\boldsymbol{R}\right)\right)
\end{align}
in which the function on the right-hand side, $U\left(\rho, S_{\text{ex}}\right)$, is the thermodynamic average potential energy of the state point with density $\rho$ and excess entropy $S_{\text{ex}}$.

Equations (\ref{eq:excessEntropy}) and (\ref{eq:potEnergyExEntropy}) apply for any system \cite{LandauLif}, but they are particularly significant for R simple systems for which they imply that the configurational adiabats are curves of invariant structure and dynamics. To prove this, we first note that, as shown in Ref. \onlinecite{Isomorph2.0}, \eq{eq:scaling} implies that the excess-entropy function is invariant under uniform scaling, i.e., it  only depends on the reduced coordinate vector $\tbR$:

\begin{align}\label{eq:excessEntropyInv}
S_{\text{ex}}(\boldsymbol{R}) = S_{\text{ex}}(\tilde{\boldsymbol{R}}) \,.
\end{align}
The relation for the potential-energy function \eq{eq:potEnergyExEntropy} consequently becomes

\begin{align}\label{eq:potEnergyExEntropyReduced}
U\left(\boldsymbol{R}\right) = U (\rho, S_{\text{ex}}(\tilde{\boldsymbol{R}}))\,.
\end{align}
Isomorphs are defined as the configurational adiabats, i.e., curves along which the excess entropy is constant, in the region of phase diagram where the system is R simple \cite{paper4}. To demonstrate invariance of structure and dynamics along the isomorphs we show that Newton's second law in reduced units is invariant along an isomorph. In complete generality, this law is in reduced units (note that the particle mass is absorbed into the reduced time)

\begin{align}\label{eq:NsecondLaw}
\frac{d^{2} \tilde{\boldsymbol{R}}}{d\tilde{t}^{2}} 
= \frac{\boldsymbol{F}\left(\boldsymbol{R}\right)}{\rho^{1/d}\,k_BT}
\equiv \tilde{\boldsymbol{F}}({\boldsymbol{R}}) \,.
\end{align}
For R simple systems \eq{eq:potEnergyExEntropyReduced} implies

\begin{align}
\boldsymbol{F} = -\nabla U 
= - \left( \frac{\partial U}{\partial S_{\text{ex}}} \right)_{\rho} 
\nabla S_{\text{ex}}(\tilde{\boldsymbol{R}}) \,.
\end{align}
Using $\nabla = \rho^{1/d}\tilde{\nabla}$ and $\left(\partial U/\partial S_{\text{ex}}\right)_{\rho}=T$, the above expression becomes

\begin{align}
\boldsymbol{F} = - \rho^{1/d}T\tilde{\nabla}S_{\text{ex}}(\tilde{\boldsymbol{R}})
\end{align}
or

\begin{align}
\tilde{\boldsymbol{F}} = - \frac{\tilde{\nabla}S_{\text{ex}}(\tilde{\boldsymbol{R}})}{k_{B}}  \,.
\end{align}
This expression reveals that for R simple systems the reduced force vector $\tilde{\boldsymbol{F}}$ is a function of the reduced coordinate vector, implying that \eq{eq:NsecondLaw} is invariant along an isomorph. Thus the dynamics is isomorph invariant in reduced units, which implies that the reduced-unit structure is also isomorph invariant \cite{Isomorph2.0}.

\subsection{The density-scaling exponent}

When strong correlations between virial and potential energy are present, a constant of proportionality between the instantaneous fluctuations of these two quantities' deviation from their equilibrium values, $\gamma$, can be introduced via

\begin{align}\label{eq:propConst}
\Delta W \left(t\right) \simeq \gamma \Delta U \left(t\right) \,.	
\end{align}
The correlation coefficient associated with this linear regression is that given in \eq{eq:correlation}. The exact definition of $\gamma$ is the following \cite{paper4}:

\begin{align}\label{eq:gammaDefinition}
\gamma = \frac{\langle\Delta W\Delta U\rangle}{\langle \left(\Delta U\right)^{2}\rangle} \,.
\end{align}
By applying a standard fluctuation relation and the volume-temperature Maxwell relation one can show \cite{paper4} that

\begin{align}\label{eq:gammaIsoDef}
\gamma = \left(\frac{\partial \ln T}{\partial \ln \rho}\right)_{S_{\text{ex}}}\,.
\end{align}

The number $\gamma$ is termed the density-scaling exponent \cite{rol05,paper4}. It determines the configurational adiabats, which as mentioned are isomorphs in the R simple region of the phase diagram. If variations of $\gamma$ are insignificant, these curves are via \eq{eq:gammaIsoDef} given by $\rho^\gamma/T=$ Const. In the general case, \eq{eq:gammaIsoDef} can be used to trace out isomorphs step-by-step by repeatedly changing density by typically an amount of order 1\%, calculating the temperature change via \eq{eq:gammaIsoDef}, recalculating $\gamma$ from \eq{eq:gammaDefinition} at the new state point, etc.

In early publications \cite{paper1,paper2} we identified the constant of proportionality $\gamma$ of \eq{eq:propConst} by the following symmetric fluctuation expression, which was later renamed $\gamma_2$ to distinguish it from $\gamma$ \cite{paper4}:

\be\label{gamma2}
\gamma_2 = \sqrt{\frac{\langle(\Delta W)^2\rangle}{\langle \left(\Delta U\right)^{2}\rangle}}\,.
\ee
Whenever the correlation coefficient $R$ is close to unity, one has $\gamma \cong \gamma_2$ since the following applies

\be\label{gammagammarel}
\gamma =\gamma_2\, R\,.
\ee

Now that the main ingredients of the isomorph theory have been introduced, we proceed to present the simulation results.

\section{Results and discussion}\label{sec:results}

\subsection{Generalized Lennard-Jones pair potentials in three dimensions}

If $r$ is the interparticle distance, the generalized LJ pair potential is defined as follows

\begin{align}\label{eq:genLJ}
v_{m,n}^{\text{LJ}} \left( r \right) = \frac{\varepsilon}{m - n} \left[ n \left( \frac{\sigma}{r} \right)^{m} - m \left( \frac{\sigma}{r} \right)^{n} \right]\,.
\end{align}
Here $m$ and $n$ are positive integers, and in order to ensure thermodynamic stability it is assumed that $m > n$. The constants $\sigma$ and $\varepsilon$ define the potential's length and energy scales, respectively, and the normalization used in \eq{eq:genLJ} ensures that the minimum pair potential energy is $-\varepsilon$ which is obtained at $r=\sigma$. Note that the normalization is different from that usually employed for the standard 12-6 LJ pair potential parametrized as $4\varepsilon\left[(r/\sigma)^{-12}-(r/\sigma)^{-6}\right]$.

\begin{figure}[htbp]
\centering
\includegraphics[width=0.45\textwidth]{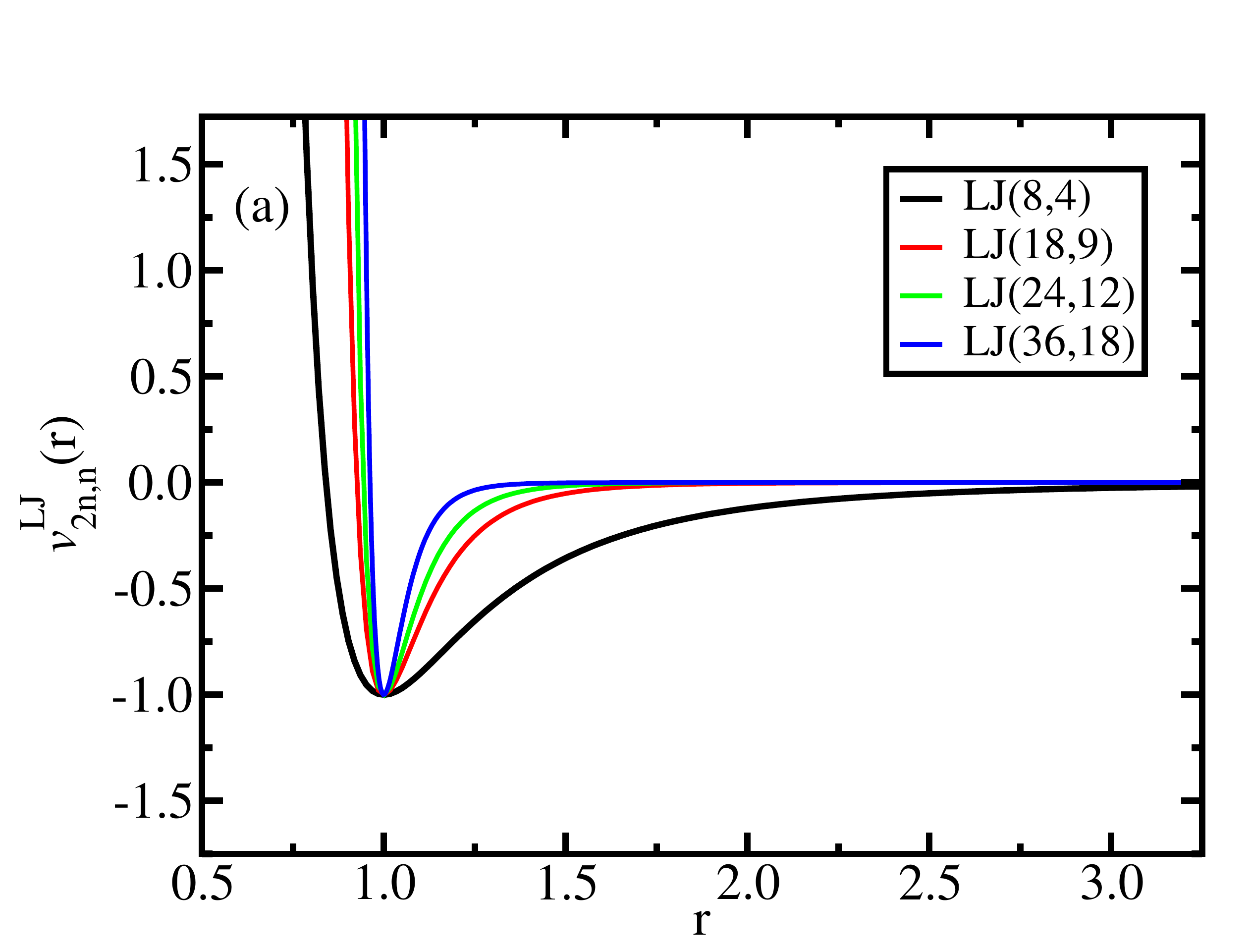}
\caption{
The simulated four generalized Lennard-Jones (LJ) pair potentials of \eq{eq:genLJcase} plotted as a function of the pairwise distance $r$ (distance and energy are given in LJ units): $n=4$ (black), $n=9$ (red), $n=12$ (green), and $n=18$ (blue). The potentials are quite different.
 } \label{fig:potentialPlot}
\end{figure}

The aim of our study is to investigate whether any relation between the correlation coefficient and the density-scaling exponent can be determined. This involves simulating several state points for which the virial potential-energy correlations are not strong. With this goal in mind, a particular case of the generalized LJ potential was simulated, the case where $m = 2n$:

\begin{align}\label{eq:genLJcase}
v_{2n,n}^{\text{LJ}} \left( r \right) = \varepsilon \left[ \left( \frac{\sigma}{r} \right)^{2n} - 2 \left( \frac{\sigma}{r} \right)^{n} \right] \text{.}
\end{align}
The following four instances of this pair potential, henceforth denoted by LJ(2n,n), were simulated in three dimensions: $n = 4, n = 9, n = 12$ and $n = 18$ (the $n=6$ case corresponding to the standard LJ potential is considered later). Figure \ref{fig:potentialPlot} shows that these four pair potentials are quite different as functions of the pairwise distance $r$.

The simulations were performed in the $NVT$ ensemble, i.e., for a constant number of particles ($N=864$ for $n=4,9,12$ and $N=4000$ for $n=18$) at constant temperature $T$ and constant volume $V$. The time step was $0.001$ in LJ units and the simulations were performed with periodic boundary conditions and a standard shifted potential cut-off at $2.5\sigma$. Using an FCC crystal as starting configuration, each system was equilibrated for $2 \cdot 10^{6}$ time steps before the collection of data began. After equilibration, the simulation ran for $50 \cdot 10^{6}$ time steps during which data were collected. For each of the four systems, six densities were considered, $\rho = 0.25, 0.50, 0.75, 1.00, 1.25, 1.50$ (LJ units); the temperature was varied within the range $0.25<T<5.00$ (LJ units). At each state point the correlation coefficient and the density-scaling exponent were calculated from \eq{eq:correlation} and \eq{eq:gammaDefinition}, respectively.

What to expect for the behavior of $R$ when density is lowered? To answer this question we refer to the isomorph theory, which works well for the LJ and related pair potentials in the ordinary liquid phase. According to isomorph theory, if a reference state point $(\rho_0,T_0)$ and another state point $(\rho,T)$ are on the same isomorph, the ratio between $T$ and $T_{0}$ defines what may be termed the isomorph shape function $h\left(\rho, S_{\text{ex}}\right)$ via $T/T_0=h\left(\rho, S_{\text{ex}}\right)$ \cite{Lasse2012}. Thus each isomorph is mapped out in the phase diagram by 

\be\label{hTrel}
\frac{h\left(\rho, S_{\text{ex}}\right)}{T}\,=\,{\rm Const.}\
\ee
The original isomorph theory predicted the function $h\left(\rho, S_{\text{ex}}\right)$ to be independent of $S_{\text{ex}}$, whereas the more correct theory from 2014 predicts that $h$ may vary slightly from isomorph to isomorph \cite{Isomorph2.0}. In both cases, however, the analytical form of the density dependence is the same. For the $\text{LJ}(2n, n)$ pair potential it can be shown that the isomorph shape function takes the following form in $d$ dimensions \cite{Lasse2012, hrhoTrond, PhDthesis,Isomorph2.0}

\begin{align}\label{eq:shapeFunction}
h\left(\rho, S_{\text{ex}}\right)
= \left(\frac{d}{n}\gamma_0 - 1 \right) \left(\frac{\rho}{\rho_{0}}\right)^{2n/d} - \left(\frac{d}{n}\gamma_0 - 2 \right) \left(\frac{\rho}{\rho_{0}}\right)^{n/d}\,.
\end{align}
Here $(\rho_0, T_0)$ is the isomorph's reference state point, at which $\gamma_0$ is the density-scaling exponent -- the $\Sex$ (isomorph) dependence of the function $h$ is contained in $\gamma_0$. Combining Eqs. (\ref{hTrel}) and (\ref{eq:shapeFunction}) one can map out an isomorph from information obtained by computer simulations at the reference state point, i.e., with no need to use \eq{eq:gammaIsoDef} in a tedious step-by-step process. 

Note that \eq{hTrel} and the definition of the density-scaling exponent \eq{eq:gammaIsoDef} implies \cite{ing12a}

\be\label{ghr}
\gamma\,=\,\frac{d\ln h}{d\ln\rho}\,.
\ee
Thus when the density-scaling exponent is known at a single reference state point, the theory via \eq{eq:shapeFunction} and \eq{ghr} predicts how $\gamma$ varies with density along the isomorph in question.

In \eq{eq:shapeFunction} there are two terms, one positive and one negative. At high densities the positive term dominates. Upon lowering the density a point is reached where $h(\rho, S_{\text{ex}})$ changes sign. Below this density the theory implies negative isomorph temperatures, compare \eq{hTrel}, which shows that the isomorph theory must break down here. Since the theory works well whenever there are strong virial potential-energy correlations, one concludes that the correlation coefficient must decrease upon lowering density along a configurational adiabat, i.e., for generalized LJ systems in any dimension the isomorph theory predicts its own breakdown at sufficiently low densities.

\begin{figure}[htbp]
	\includegraphics[width=7cm]{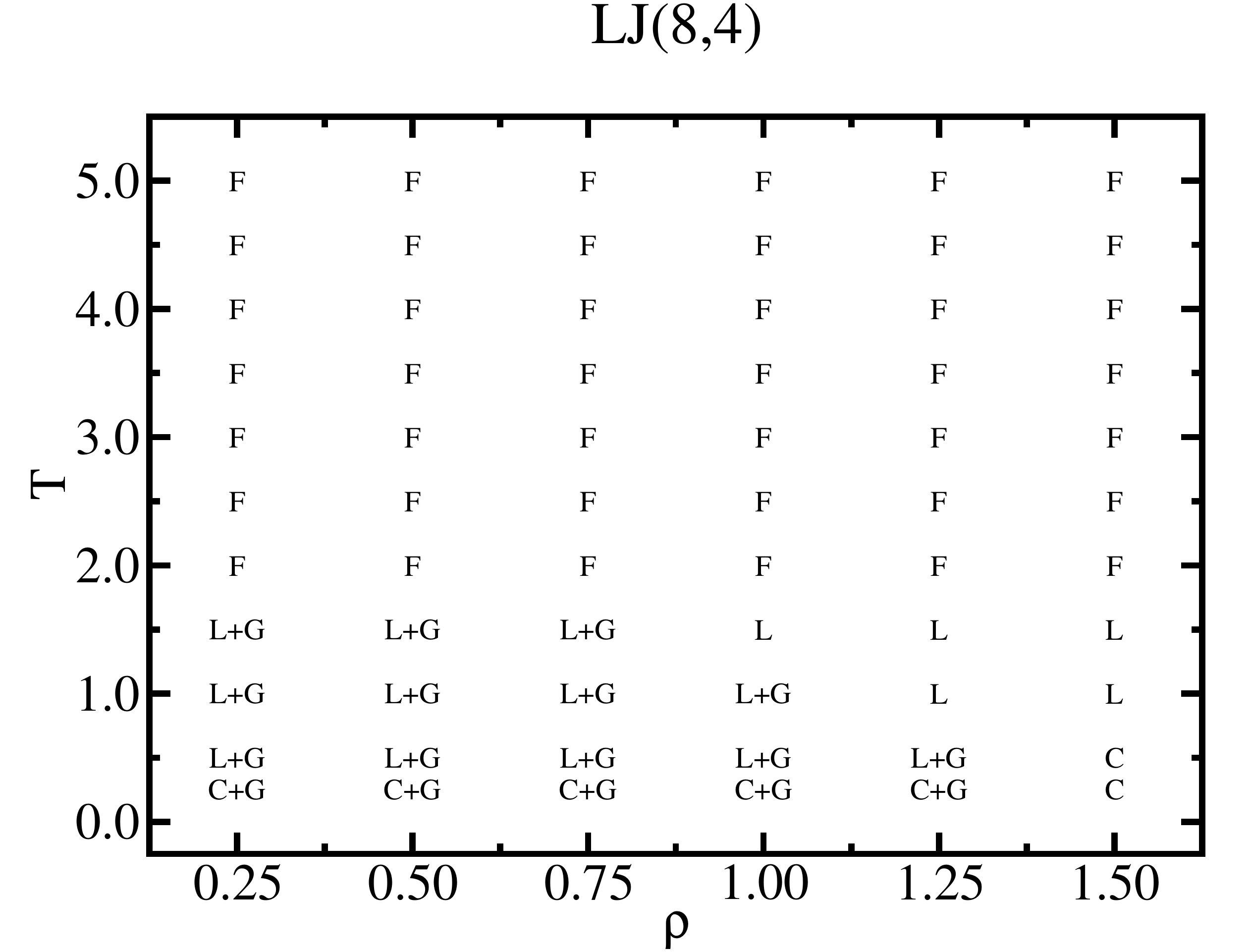}
	\includegraphics[width=7cm]{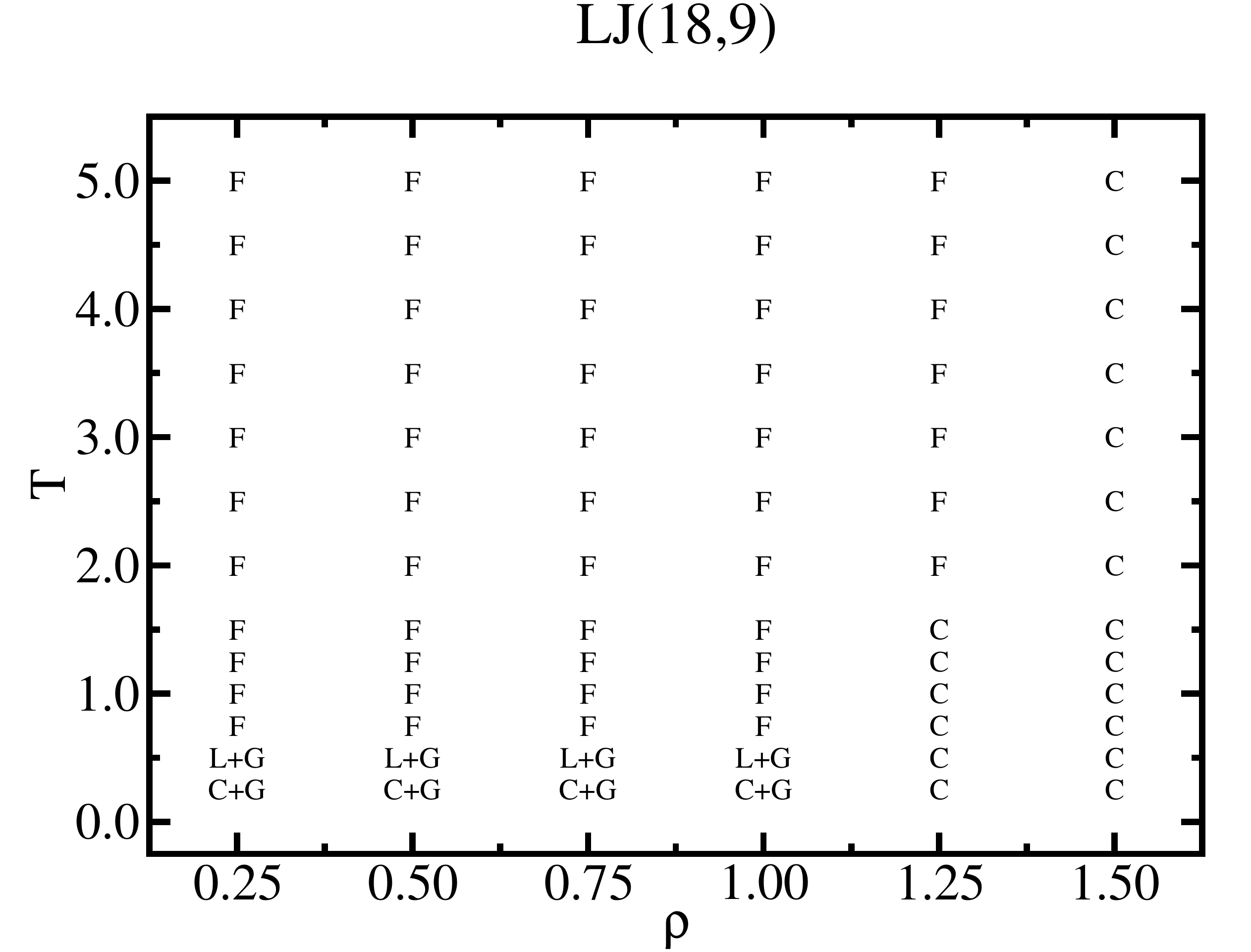} 
	\includegraphics[width=7cm]{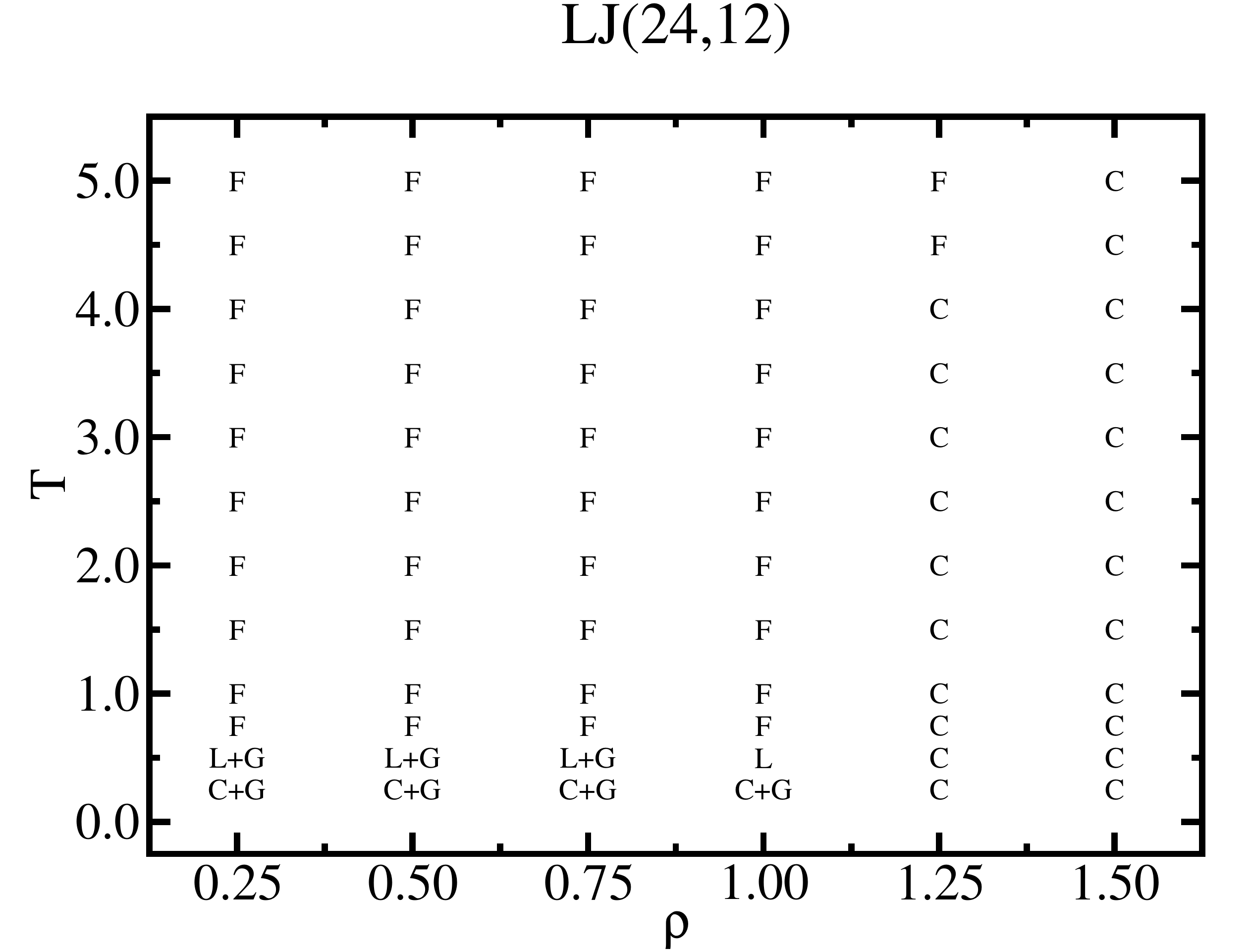} 
	\includegraphics[width=7cm]{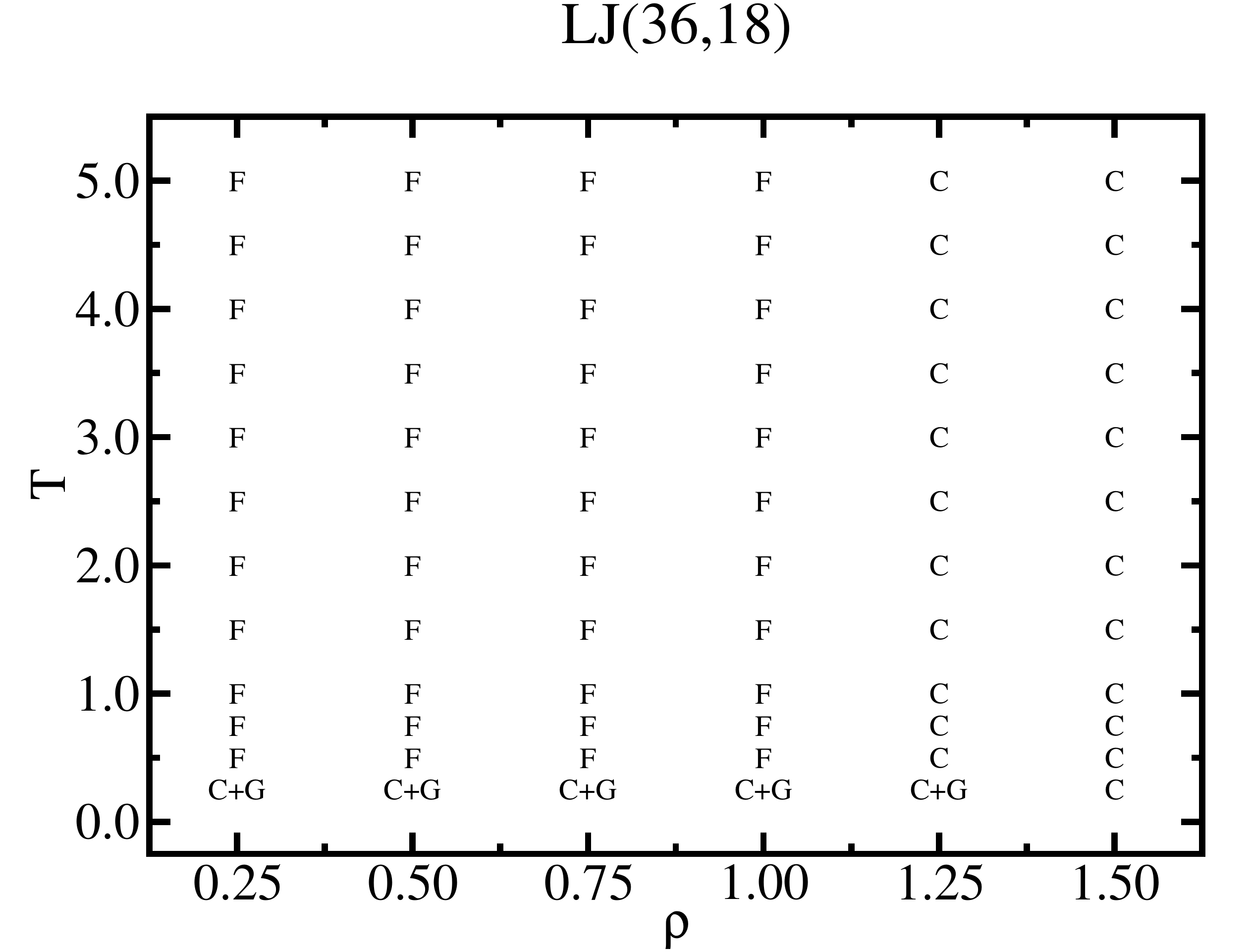}
	\caption{
		State points simulated for the pair potentials LJ(8,4), LJ(18,9), LJ(24,12), LJ(36,18) defined by \eq{eq:genLJcase}.  At each state point the virial potential-energy correlation coefficient (\eq{eq:correlation}) and the density-scaling exponent (\eq{eq:gammaDefinition}) were evaluated. Most state points are at supercritical temperatures (at which the liquid and gas phases merge) and marked by an F (F means fluid, L means liquid, C means crystalline, L+G means a state point of coexisting liquid and gas phases, etc). The phases were identified by visual inspection of selected configurations.
		}\label{fig:Phase_diagram}
\end{figure}

To summarize, moving along a given isomorph it is possible to define a density below which the isomorph theory does not hold. The condition that the shape function is non-negative sets the following limit for the isomorph theory to work:

\be
\rho\,>\, \left(\frac{d \gamma_0 - 2n}{d \gamma_0 - n}\right)^{d/n} \rho_{0}\,.
\ee

We have seen that $R$ decreases as density is lowered along a configurational adiabat (isomorph). Moreover, according to \eq{eq:gammaDefinition} the scaling exponent must approach zero if $R\rightarrow 0$.  Since $R=0$ implies $\gamma=0$, one expects that $\gamma$ also decreases upon lowering the density. If density is increased, on the other hand, the positive term of \eq{eq:shapeFunction} dominates and the density-scaling exponent eventually approaches $2n/d$ due to the dominance of the repulsive $r^{-2n}$ term \cite{Lasse2012, hrhoTrond, PhDthesis}. At the same time $R$ tends to unity. After considering these two limits, we turn to the simulation results.

\begin{figure}[htbp]
  \includegraphics[width=7cm]{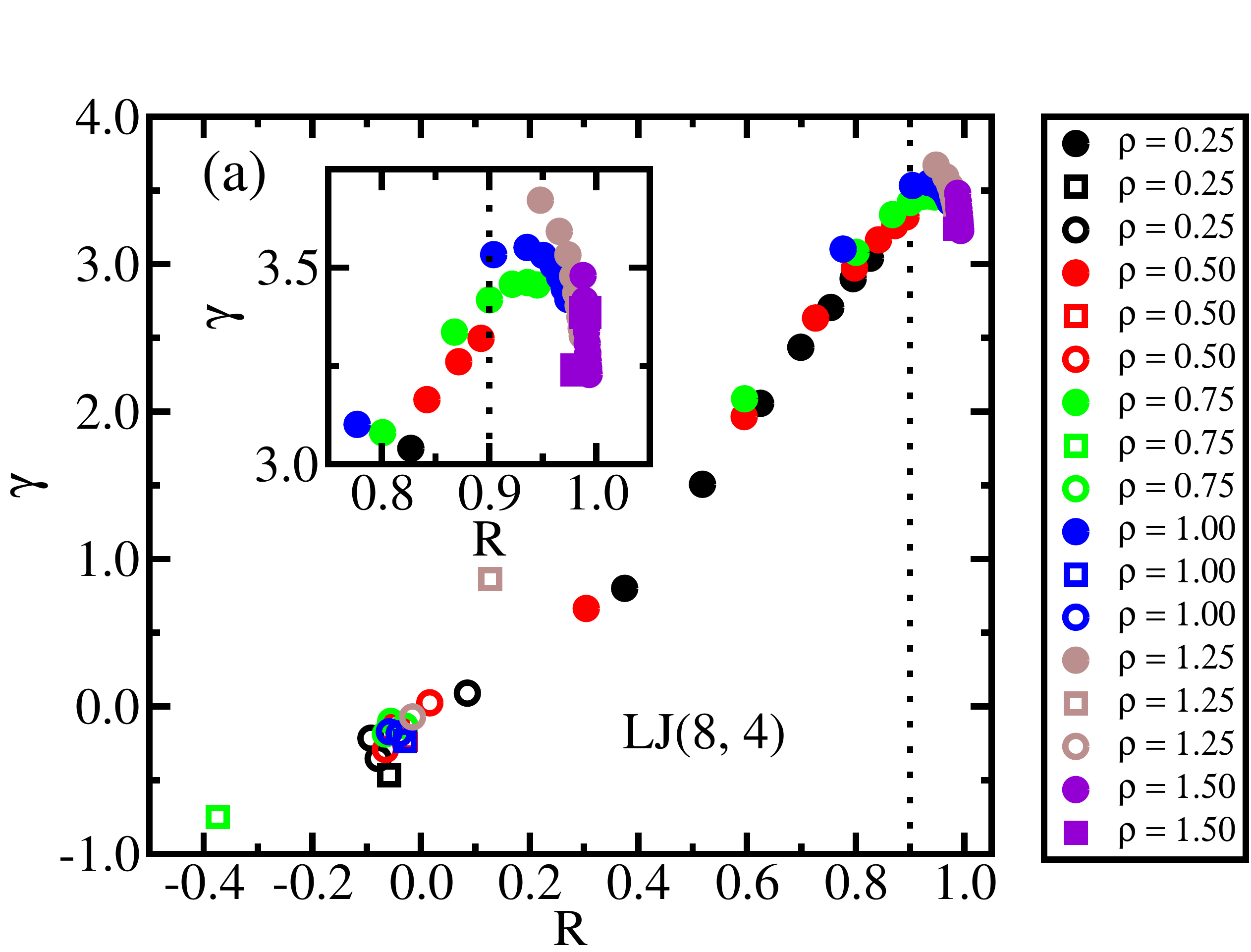}
  \includegraphics[width=7cm]{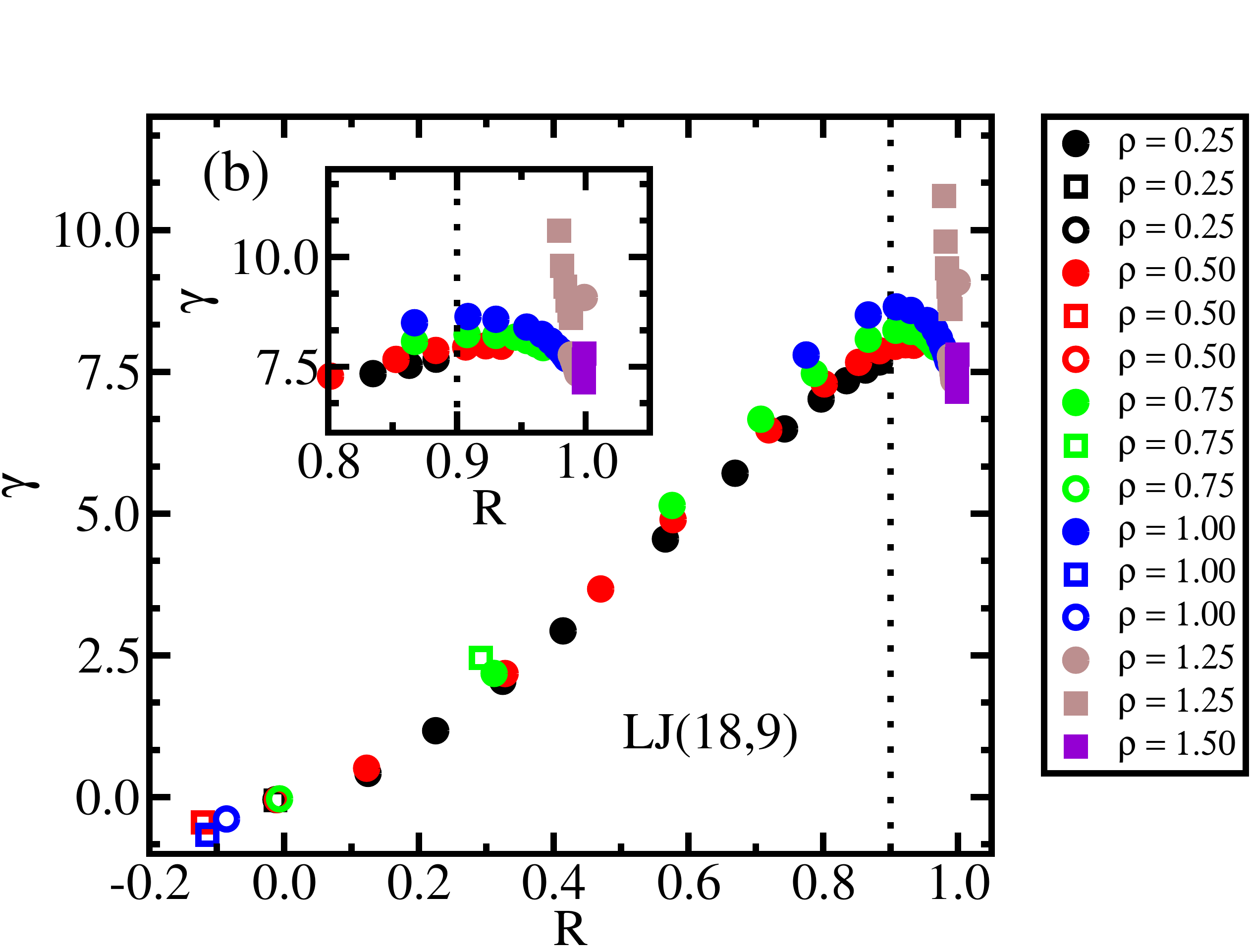} 
  \includegraphics[width=7cm]{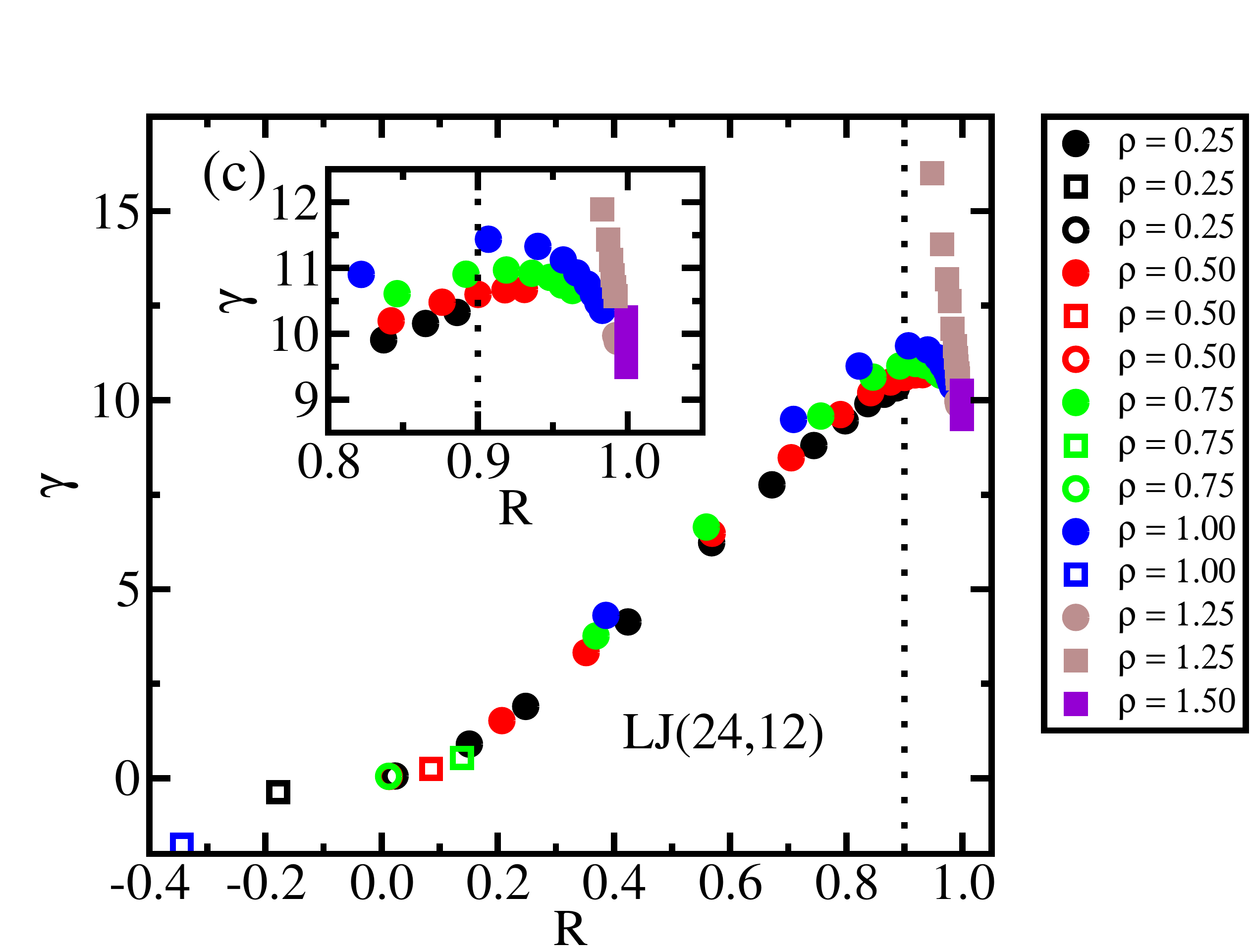} 
  \includegraphics[width=7cm]{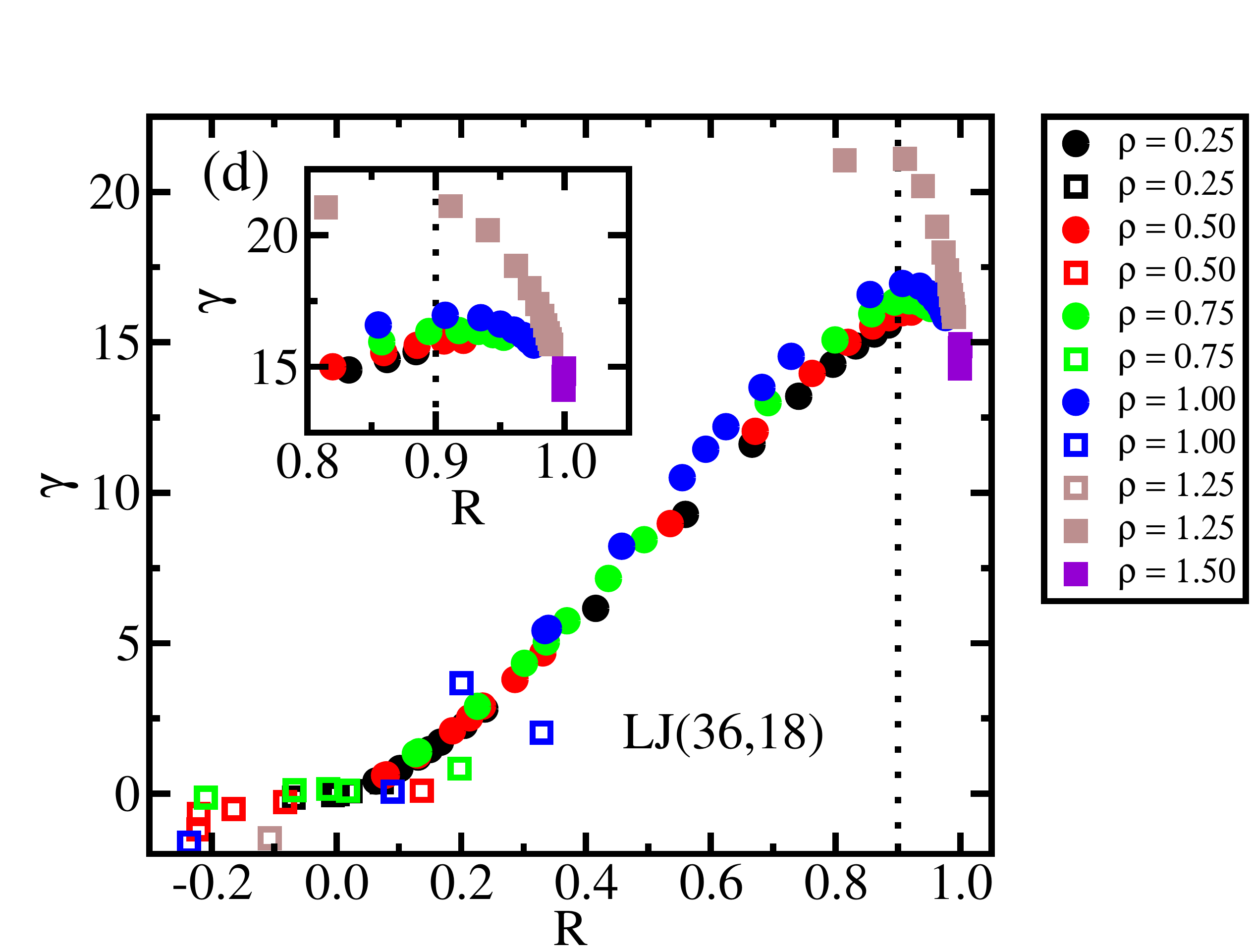}
  \caption{
The density-scaling exponent, $\gamma$, versus the correlation coefficient, $R$, for generalized LJ potentials: 
(a) LJ(8,4); (b) LJ(18,9); (c) LJ(24,12); (d) LJ(36,18). 
Full squares mark crystalline states, full circles are liquid, fluid, or gas states -- open squares are coexisting crystal-gas states; open circles are coexisting liquid-gas states. The simulated state points are given in \fig{fig:Phase_diagram}. The inset in each figure focuses on the crossover region. Along each isochore the temperature was varied in the range $0.25<T< 5.00$. In (b), (c), and (d) it is possible to distinguish two different behaviors of the variation of $\gamma$ with $R$, which reflects liquid contra crystal phases (the latter being the steepest). The state points with negative or close to zero correlation coefficient are all in the gas-liquid or gas-solid coexistence regions. }\label{fig:gammaR}
\end{figure}

Figure \ref{fig:Phase_diagram} shows the state points studied, marking in each case whether it is a solid or liquid state point, etc. Most state points are supercritical in which case we marked them by an F indicating ``fluid''. Figure \ref{fig:gammaR} plots $\gamma$ versus $R$ along the isochores for the $\text{LJ}(2n, n)$ potentials with $n=4,9,12,18$. Our findings are consistent with the theoretically predicted result that as $R\rightarrow 1$, $\gamma$ approaches the high-density limiting values 8/3, 6, 8, and 12, respectively, although we never come really close to these values that apply whenever the attractive term of the potential may be completely ignored. The density-scaling exponent changes behavior and starts to decrease with increasing $R$ in the region where $R$ is around $0.9$ which, as mentioned earlier, is the threshold usually used to define R simple systems. In the opposite limit, $R\rightarrow 0$, we find $\gamma\rightarrow 0$, as expected. 
 
Our data show that the change of behavior in $\gamma$ versus $R$ takes place at the boundary of the region in which the system is R simple. A more extensive study is needed to determine whether this holds also for other R simple systems. If confirmed, the relation between the change in the behavior of the density-scaling exponent and the value of the correlation coefficient could be useful in practice, because $R$ is not accessible in experiment while $\gamma$ is \cite{Gundermann2011}. A change in the density or temperature dependence of $\gamma$ might therefore be used for identifying in which regions of the phase diagram a liquid is expected to obey isomorph theory predictions like isochronal superposition, excess entropy scaling, etc \cite{hiddenscale}.

\begin{figure}[htbp]
	\centering
  \resizebox{\textwidth}{!}{
  \begin{tabular}{l l}
  \includegraphics[width=7cm]{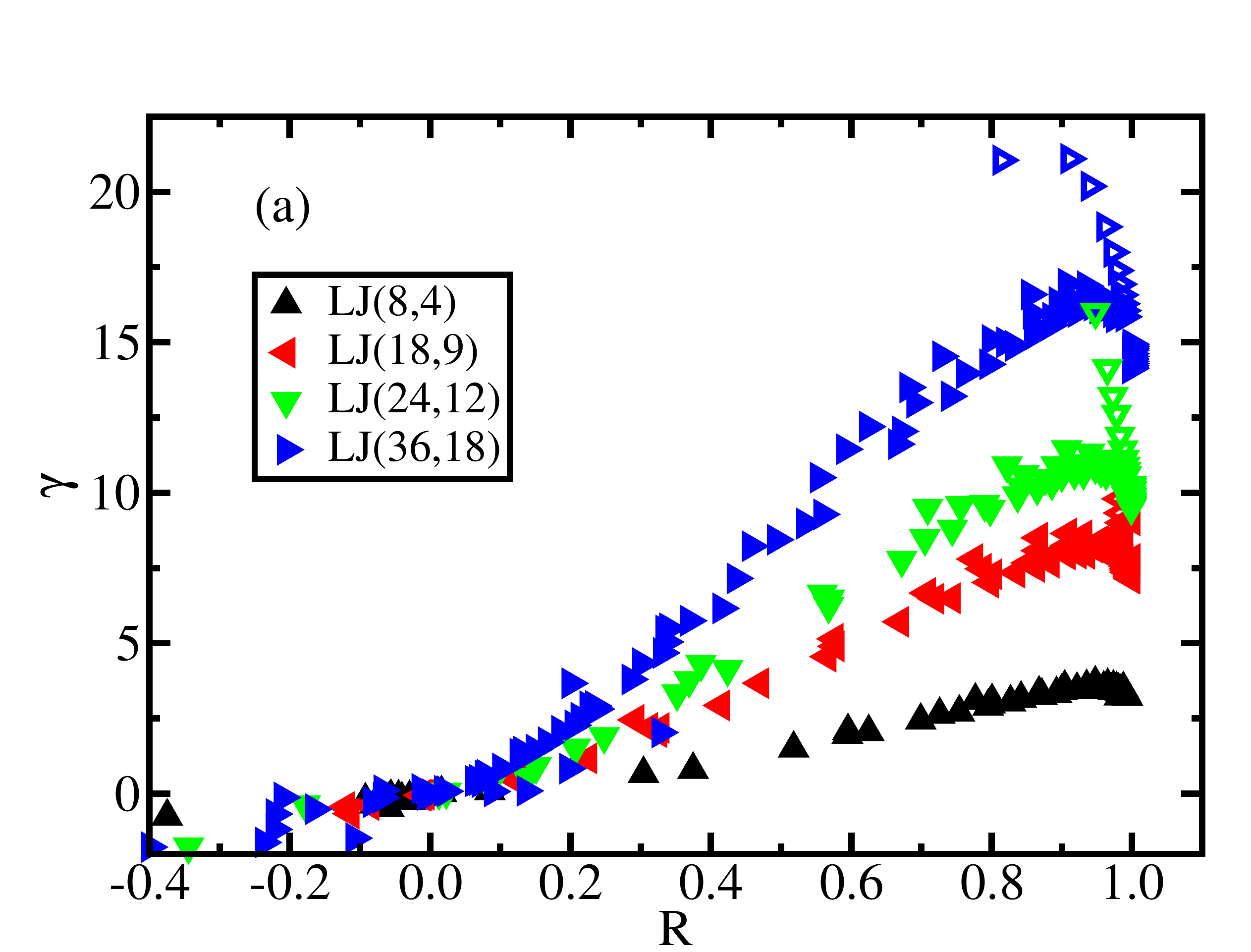} &
  \includegraphics[width=7cm]{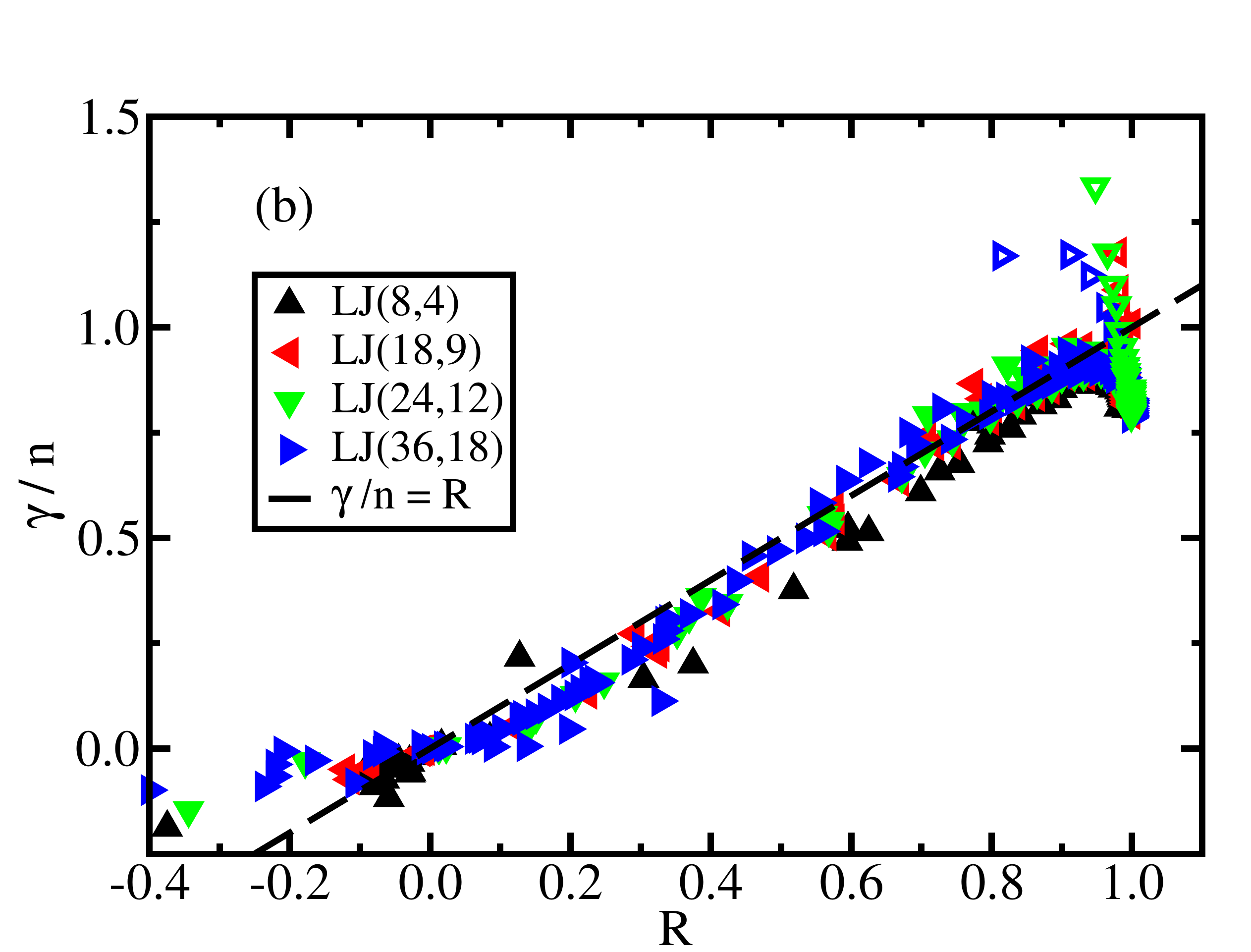} \\
  \end{tabular}}
  \caption{(a) All data of \fig{fig:gammaR} in a common plot in which open triangles are crystalline states and full triangles are liquid, fluid, gas, or coexistence states. 
  	(b) The same data with $\gamma/n$ plotted versus $R$. The black dashed line connects the points $(0,0)$ with $(1,1)$, demonstrating the universal, roughly linear dependence of \eq{gnR}. Significant deviations from this are seen mainly for $R>0.9$ and $R<0$. The former case refers to the R simple region of the phase diagram, the latter to states of coexisting liquid and gas phases (compare to \fig{fig:Phase_diagram}).
  	}\label{fig:Rvsgamma}
\end{figure}

Figure \ref{fig:Rvsgamma} summarizes the data of \fig{fig:gammaR}. There is a roughly linear behavior between $\gamma$ and $R$, but it does not extend above $R\approx 0.9$ where $\gamma$ as mentioned starts to decrease towards the value $2n/3$ predicted at high density (where the $r^{-2n}$ repulsive term dominates and the virial potential-energy correlations become virtually perfect). For $R>0.9$ the system is R simple and the density variation of $\gamma$ is well described by Eqs. (\ref{eq:shapeFunction}) and (\ref{ghr}). For this reason we henceforth focus on the region in which $R<0.9$.

Figure \ref{fig:Rvsgamma}(b) plots the data differently and establishes for most of the $R<0.9$ data the following approximate proportionality:

\be\label{gnR}
\gamma/n \simeq R\,.
\ee
How to understand this? In Ref. \onlinecite{paper2} it was shown that an LJ-type pair potential may be approximated by the ``extended inverse power law (eIPL)'' pair potential with an effective state-point-dependent exponent $p$ that is \textit{not} simply the exponent of the repulsive term of the LJ-type potential:

\be\label{eIPL}
v(r)
\,\cong\, Ar^{-p}+B+Cr\,.
\ee
This approximation usually works very well within the entire first coordination shell \cite{paper2}. At typical condensed-matter low and moderate pressure state points the term $B+Cr$ contributes little to the forces on a given particle, because the sum of these terms over all nearest neighbors tends to be almost constant. This reflects the fact that if the particle is moved slightly, some nearest-neighbor distances increase and some decrease, but their sum stays almost constant \cite{paper2}. This implies that the total force contribution from the $B+Cr$ term is small and the physics is dominated by the $r^{-p}$ IPL term. 

For an IPL pair potential $\propto r^{-p}$ the density-scaling exponent is given in $d$ dimensions by \cite{paper2} 

\be\label{dsp}
\gamma\,=\,\frac{p}{d}\,.
\ee
As mentioned, the effective exponent $p$ of \eq{eIPL} varies with state point. In large parts of the phase diagram of the LJ(m,n) pair potential the effective exponent $p$ is considerably larger than $m$ \cite{paper2}. This is because even within the repulsive \textit{part} of the pair potential, i.e., at distances below the potential minimum, there is a sizable contribution from the attractive $r^{-n}$ \textit{term} making the repulsions significantly steeper than expected from the $r^{-m}$ repulsive term alone. At densities and temperatures dominated by the pair potential minimum, i.e., at typical low or moderate pressure condensed-matter state points, it may be shown from a curvature argument referring to the potential-energy minimum that the effective exponent of the generalized LJ pair potential \eq{eq:genLJ} is given \cite{paper2} by

\be\label{gap0}
p\,=\,m+n\,.
\ee
At sufficiently high density and/or temperature the repulsive $r^{-m}$ term dominates the physics, however, and here one finds that $p\rightarrow m$. These conditions arise as $R\rightarrow 1$. 

For the standard LJ pair potential at typical state points the density-scaling exponent is between $5$ and $6$, corresponding via \eq{dsp} to $p$ between $15$ and $18$ \cite{paper1,paper2}. Based on the above, we expect that the LJ(2n,n) pair potential in a large part of its phase diagram is equivalent to an eIPL pair potential with $p\simeq 2n+n=3n$. Via \eq{dsp}, this corresponds in three dimensions to

\be\label{gap}
\gamma\simeq n\,.
\ee

In summary, $R\rightarrow 1$ at state points of high density and/or temperature where the $r^{-2n}$ term dominates the physics. This behavior is hinted at in \fig{fig:Rvsgamma}(b) above $R\simeq 0.9$ where $\gamma/n$ starts to decrease and systematically deviates from the black dashed line. The limit $\gamma\simeq 2n/3$ was not reached in our simulations, however, which shows that only at \textit{very} high density or temperature the attractive term of the generalized LJ pair potential may be ignored. A linear extrapolation of most of the data to $R=1$ gives the limit value suggested by the eIPL approximation in conjunction with  \eq{gap0}. We take this as an indication that the eIPL approximation works well at the state points simulated, whether or not these are characterized by strong virial potential-energy correlations. 

Having identified the limits $\gamma\rightarrow 0$ for $R\rightarrow 0$ and $\gamma\rightarrow n$ for $R\rightarrow 1$ (as long as $R<0.9$, i.e., for most data), the next question one may ask is: why do the data follow the approximate linear relation \eq{gnR}? A possible explanation refers again to the eIPL approximation. The virial is calculated in $d$ dimensions from the pair potential as a sum over terms of the form $rv'(r)/d$ \cite{paper1}. For the eIPL pair potential \eq{eIPL} this gives $Apr^{-p}/d$ plus an approximate constant (compare the above argument). Thus for fluctuations away from the equilibrium value, if the linear term of \eq{eIPL} is uncorrelated to the IPL term, the eIPL approximation suggests that one may have effectively

\be\label{DWpd}
\Delta W \simeq (p/d)\Delta U
\ee
with $p\cong 3n$ for $d=3$. This must be, admittedly, a very rough approximation when the $W$ $U$ correlations are not strong, but if one nevertheless makes it, \eq{eq:correlation} and \eq{eq:gammaDefinition} imply 

\be\label{gRn_rel}
\gamma/R\simeq \sqrt{\langle (\Delta W)^2\rangle} / \sqrt{\langle (\Delta U)^2\rangle}\simeq p/d
\simeq n\,.
\ee
This is \eq{gnR} that summarizes our findings for most state points in three dimensions. A concise way of expressing this is that the exponent $\gamma_2$ of \eq{gamma2} is roughly constant throughout the non-R-simple region of the phase diagram, compare \eq{gammagammarel}.

\subsection{The Lennard-Jones pair potential in two, three, and four dimensions}

\begin{figure}[htbp]
	\centering
  \resizebox{\textwidth}{!}{
  \begin{tabular}{l l}
  \includegraphics[width=7cm]{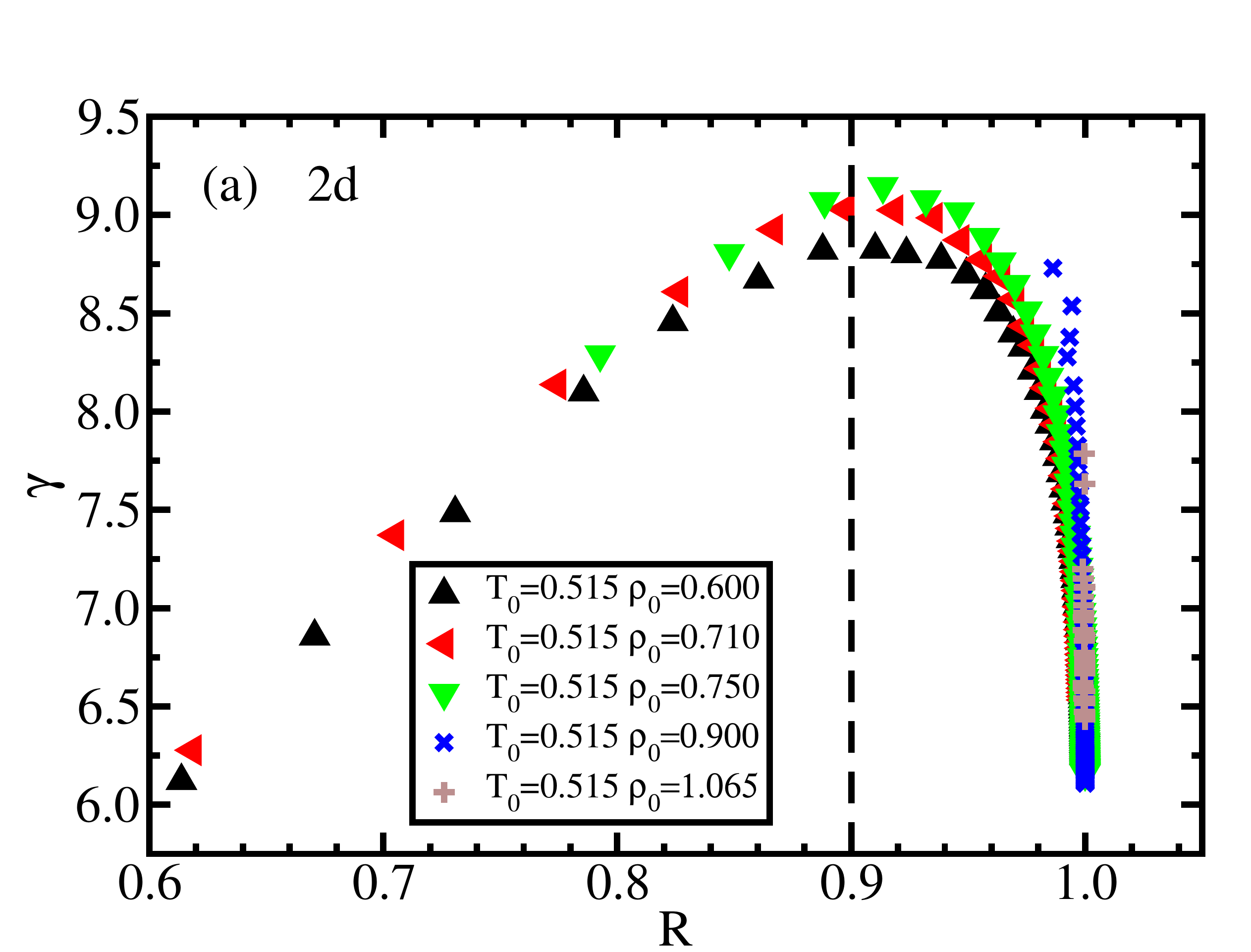} &
  \includegraphics[width=7cm]{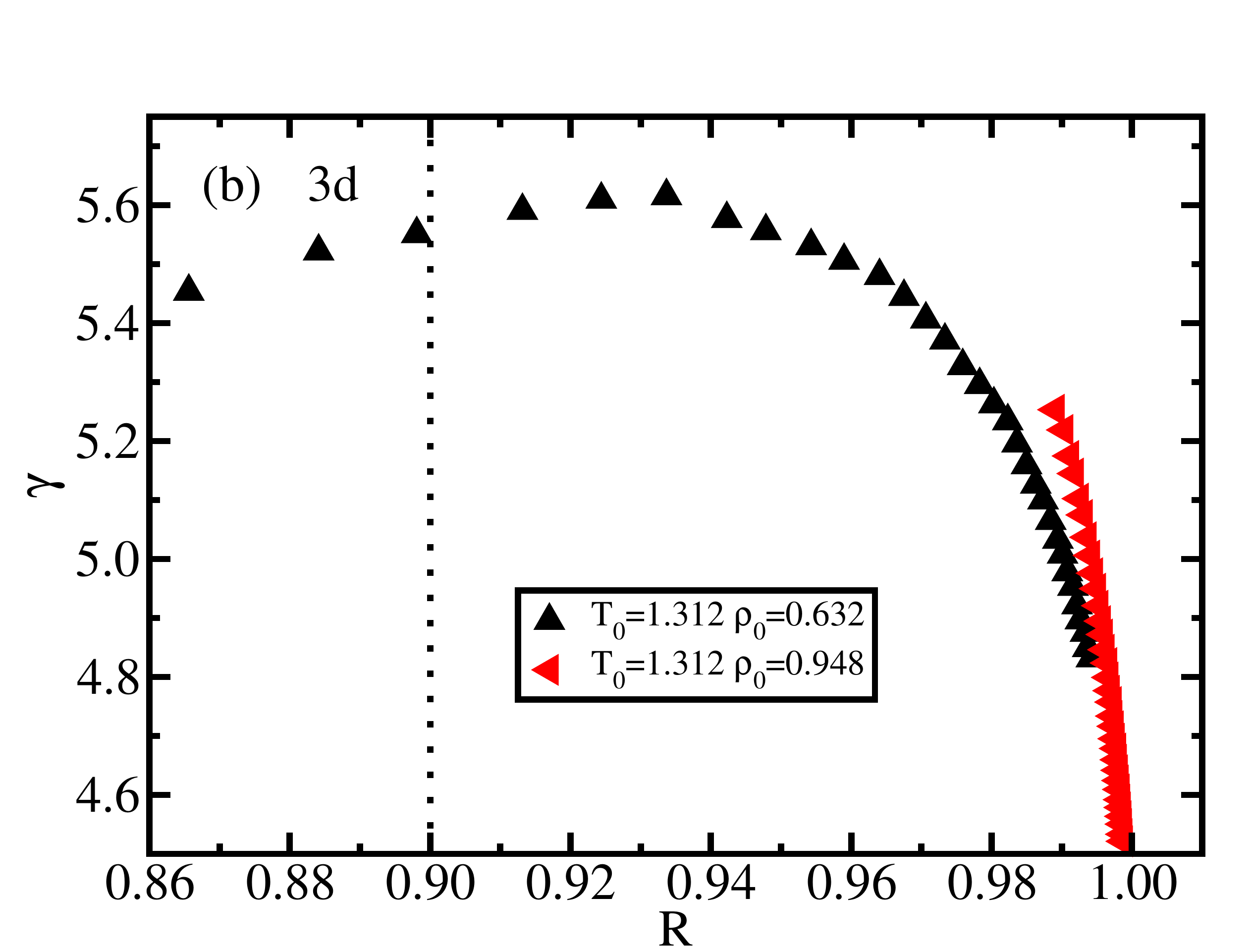}
  \end{tabular}}
  \caption{
The density-scaling exponent, $\gamma$, versus the correlation coefficient, $R$, along configurational adiabats in two and three dimensions with legends giving the reference state points. 
(a) Results for three configurational adiabats in the liquid phase and two in the crystalline phase generated using \eq{eq:gammaIsoDef} for the $2d$ LJ system. Triangles represent liquid states, crosses represent solid states. The density-scaling exponent increases with increasing $R$ for values of $R$ below the $0.9$ threshold and then starts to decrease. For the crystalline phase, $R$ is always larger than $0.9$ and $\gamma$ decreases with increasing $R$ (towards $12/2=6$). 
(b) Two configurational adiabats for the liquid phase of the $3d$ LJ system, constructed as in the $2d$ case. The decrease in $\gamma$ starts a bit after the $0.9$ threshold.}\label{fig:gammaRDim}
\end{figure}

We proceed to report results for the standard LJ system in two, three, and four dimensions. Details on how the simulations were performed can be found in Refs. \onlinecite{PhDthesis,Costigliola2016b}. Figure \ref{fig:gammaRDim}(a) shows the density-scaling exponent versus the correlation coefficient along three configurational adiabats in the liquid phase and two in the crystalline phase for the two-dimensional LJ system; (b) shows a similar plot along two configurational adiabats in the liquid phase of the three-dimensional LJ system.

\begin{figure}[htbp]
\centering
\includegraphics[width=7cm]{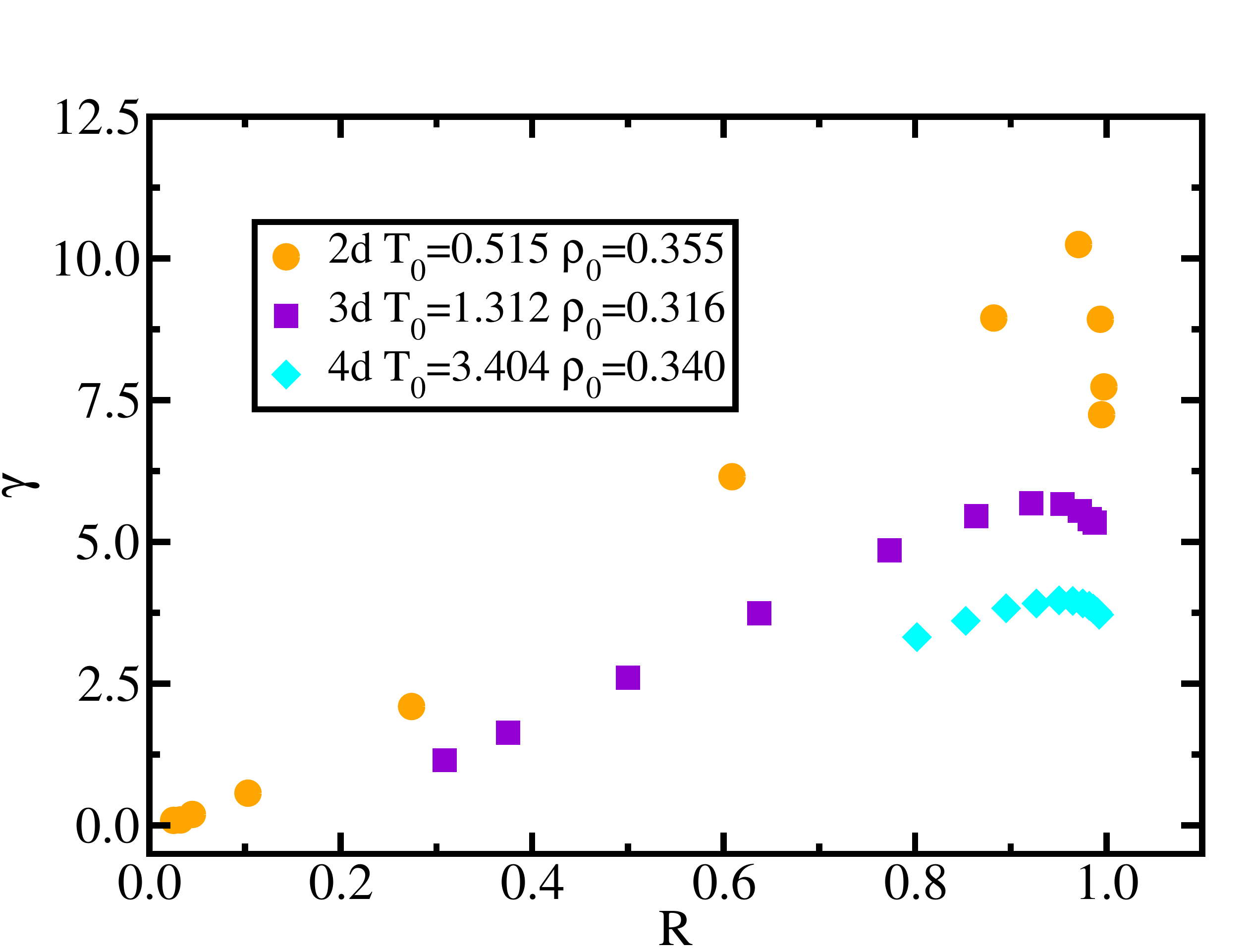}
\caption{
The density-scaling exponent, $\gamma$, versus the correlation coefficient, $R$, of the LJ system in two, three, and four dimensions along configurational adiabats using the critical points as reference state points. In all cases, $\gamma$ exhibits a maximum in the region where the correlation coefficient crosses the threshold value $R=0.9$. Note that the maximum value of $\gamma$ decreases with increasing dimension, as expected from \eq{dsp} and \eq{gap0}.
 }\label{fig:criticalisomorph}
\end{figure}

Figure \ref{fig:criticalisomorph} shows the density-scaling exponent versus the correlation coefficient along the configurational adiabats defined by the critical points for the LJ system in two, three, and four dimensions. Data for the critical points are taken from Refs. \onlinecite{Smit1991, Potoff1998, Hloucha1999}. For the configurational adiabats in the liquid phase the relation between $R$ crossing the threshold value $0.9$ and the start of the decreasing behavior of $\gamma$ is observed. For the crystalline phase, the correlation coefficient is above $0.9$ for all the state points studied and no crossover is observed.

\begin{figure}[htbp]
\includegraphics[width=7cm]{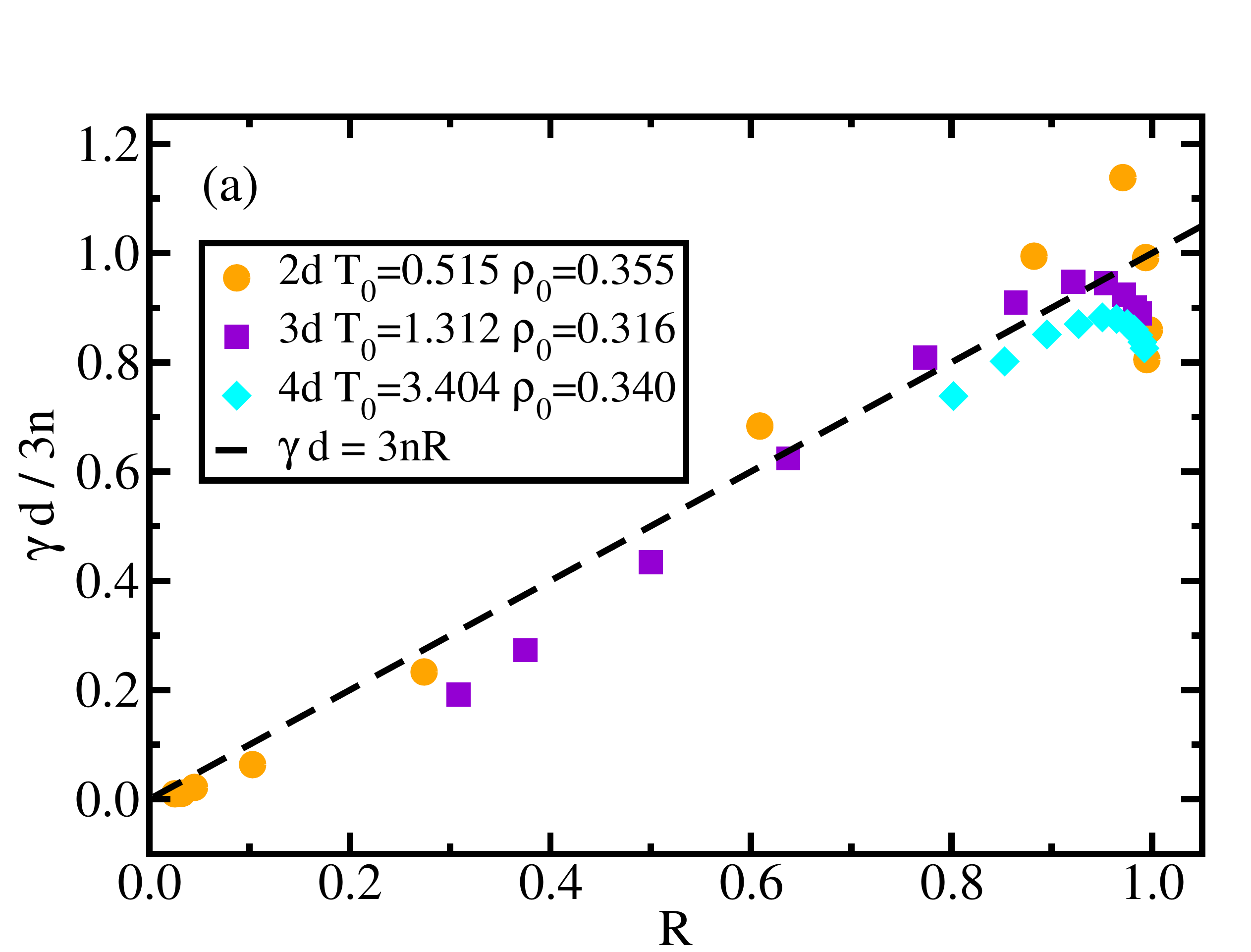}
\includegraphics[width=7cm]{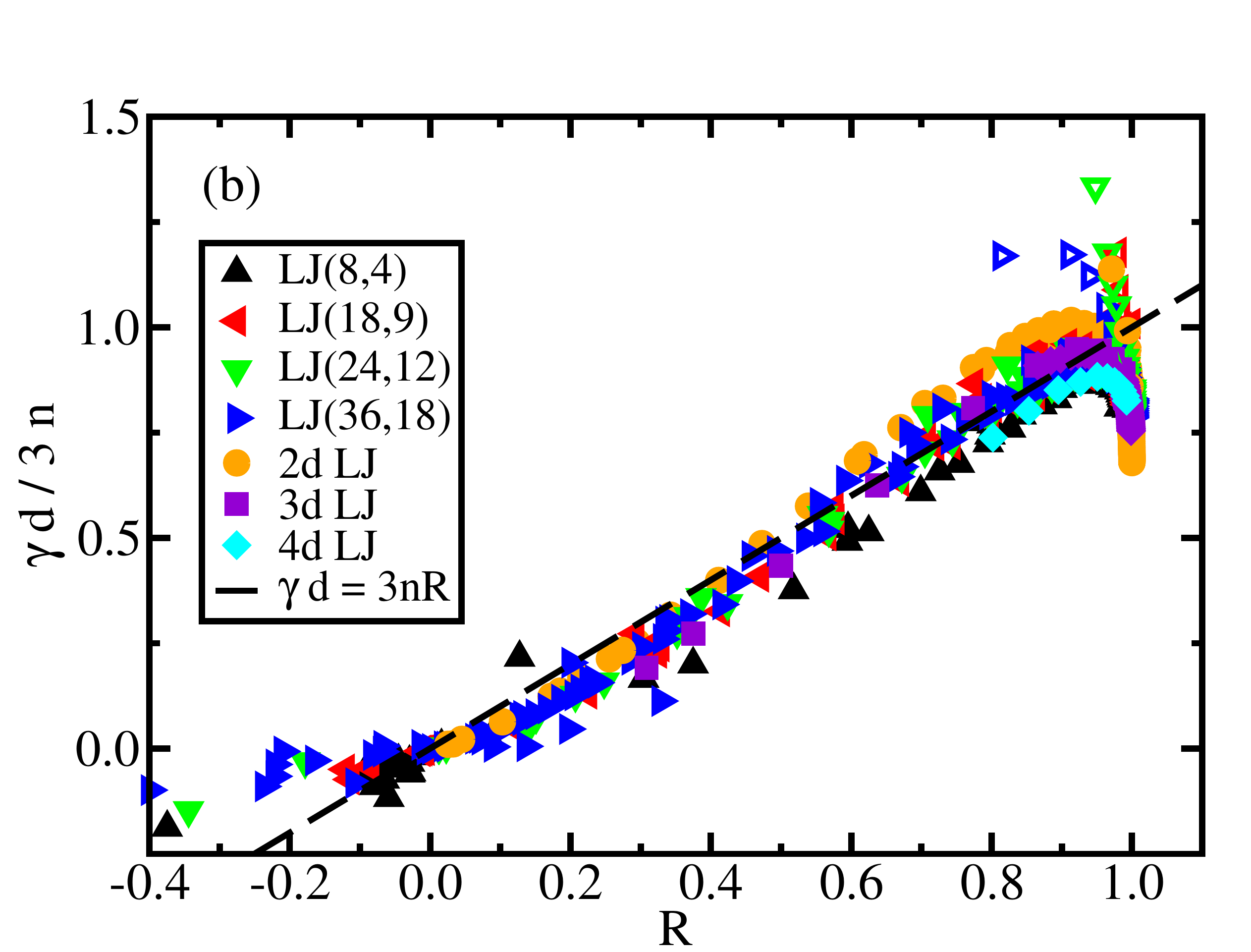}
\caption{Normalized density-scaling exponent versus the virial potential-energy correlation coefficient.
(a) The data for the LJ system in two, three, and four dimensions.
(b) All data presented in this paper, demonstrating the approximate linear relationship \eq{gnR_dim} below $R\approx 0.9$ (dashed line). Above this threshold, as discussed in the text, there is a different behavior. The open symbols represent solid states, the full symbols represent all other states (including coexistence states).
\label{gdRn_sim}
}
\end{figure}

In $d$ dimensions the eIPL-justified relation \eq{gRn_rel} implies via $p=2n+n=3n$ that

\be\label{gnR_dim}
\gamma\, d\, \cong\, 3\,n\, R\,.
\ee
This is tested in \fig{gdRn_sim}(a) for the LJ simulations in $2-4$ dimensions. We see that \eq{gnR_dim} overall works well.

Finally, \fig{gdRn_sim}(b) tests all data presented in this paper versus \eq{gnR_dim}. The data in the region where the system is not R simple conforms roughly to \eq{gnR_dim}, unless $R<0$ which are state points of liquid-gas coexistence.

\section{Conclusions}\label{sec:disc}

For the state points where the system is not R simple ($R<0.9$) the generalized Lennard-Jones systems have a roughly linear relation \eq{gnR_dim} between the density-scaling exponent and the virial potential-energy correlation coefficient of (unless $R<0$ corresponding to liquid-gas coexistence state points). This finding has been rationalized in terms of the extended inverse-power-law approximation. Above $R\cong 0.9$ the approximate linear relation $\gamma\propto R$ breaks down; when $R$ approaches unity, the density-scaling exponent starts to decrease with increasing $R$. $\gamma$ is expected to eventually obey $\gamma\rightarrow 2n/d$ as $R\rightarrow 1$ when the repulsive $r^{-2n}$ term dominates completely. This happens only at very large densities and/or temperatures, however.

Our numerical data show that the threshold value $R\cong 0.9$ has a hitherto not recognized significance. While it was previously known that this threshold delimits the state points for which the isomorph theory works well \cite{paper4,hiddenscale}, it is a new finding that roughly the same threshold marks a crossover from the $\gamma\propto R$ behavior found for $R<0.9$ to the one of decreasing $\gamma$ with increasing $R$ predicted by \eq{eq:shapeFunction} and \eq{ghr}. It would be interesting to find out how general the finding of a threshold at $R\cong 0.9$ is. If it is general, this opens for the possibility -- at least in principle since the required experiments are challenging -- that determining the experimental variation of the density-scaling exponent throughout the phase diagram may be used for determining where the system in question is R simple and where it is not.

\acknowledgments{We are indebted to Daniele Coslovich for helpful discussions regarding the interpretation of our numerical findings. This work was supported by the Danish National Research Foundation's grant DNRF61 and by the Villum Foundation's grant VKR-023455.	}


\begin{thebibliography}{31}%
\makeatletter
\providecommand \@ifxundefined [1]{%
 \@ifx{#1\undefined}
}%
\providecommand \@ifnum [1]{%
 \ifnum #1\expandafter \@firstoftwo
 \else \expandafter \@secondoftwo
 \fi
}%
\providecommand \@ifx [1]{%
 \ifx #1\expandafter \@firstoftwo
 \else \expandafter \@secondoftwo
 \fi
}%
\providecommand \natexlab [1]{#1}%
\providecommand \enquote  [1]{``#1''}%
\providecommand \bibnamefont  [1]{#1}%
\providecommand \bibfnamefont [1]{#1}%
\providecommand \citenamefont [1]{#1}%
\providecommand \href@noop [0]{\@secondoftwo}%
\providecommand \href [0]{\begingroup \@sanitize@url \@href}%
\providecommand \@href[1]{\@@startlink{#1}\@@href}%
\providecommand \@@href[1]{\endgroup#1\@@endlink}%
\providecommand \@sanitize@url [0]{\catcode `\\12\catcode `\$12\catcode
  `\&12\catcode `\#12\catcode `\^12\catcode `\_12\catcode `\%12\relax}%
\providecommand \@@startlink[1]{}%
\providecommand \@@endlink[0]{}%
\providecommand \url  [0]{\begingroup\@sanitize@url \@url }%
\providecommand \@url [1]{\endgroup\@href {#1}{\urlprefix }}%
\providecommand \urlprefix  [0]{URL }%
\providecommand \Eprint [0]{\href }%
\providecommand \doibase [0]{http://dx.doi.org/}%
\providecommand \selectlanguage [0]{\@gobble}%
\providecommand \bibinfo  [0]{\@secondoftwo}%
\providecommand \bibfield  [0]{\@secondoftwo}%
\providecommand \translation [1]{[#1]}%
\providecommand \BibitemOpen [0]{}%
\providecommand \bibitemStop [0]{}%
\providecommand \bibitemNoStop [0]{.\EOS\space}%
\providecommand \EOS [0]{\spacefactor3000\relax}%
\providecommand \BibitemShut  [1]{\csname bibitem#1\endcsname}%
\let\auto@bib@innerbib\@empty
\bibitem [{\citenamefont {{Nicholas P. Bailey and Ulf R. Pedersen and Nicoletta
  Gnan and Thomas B. Schr{\o}der and Jeppe C.
  Dyre}}(2008{\natexlab{a}})}]{paper1}%
  \BibitemOpen
  \bibfield  {author} {\bibinfo {author} {\bibnamefont {{Nicholas P. Bailey and
  Ulf R. Pedersen and Nicoletta Gnan and Thomas B. Schr{\o}der and Jeppe C.
  Dyre}}},\ }\bibfield  {title} {\enquote {\bibinfo {title} {{Pressure-energy
  correlations in liquids. I. Results from computer simulations}},}\
  }\href@noop {} {\bibfield  {journal} {\bibinfo  {journal} {Journal of
  Chemical Physics}\ }\textbf {\bibinfo {volume} {129}},\ \bibinfo {pages}
  {184507} (\bibinfo {year} {2008}{\natexlab{a}})}\BibitemShut {NoStop}%
\bibitem [{\citenamefont {{Nicholas P. Bailey and Ulf R. Pedersen and Nicoletta
  Gnan and Thomas B. Schr{\o}der and Jeppe C.
  Dyre}}(2008{\natexlab{b}})}]{paper2}%
  \BibitemOpen
  \bibfield  {author} {\bibinfo {author} {\bibnamefont {{Nicholas P. Bailey and
  Ulf R. Pedersen and Nicoletta Gnan and Thomas B. Schr{\o}der and Jeppe C.
  Dyre}}},\ }\bibfield  {title} {\enquote {\bibinfo {title} {{Pressure-energy
  correlations in liquids. II. Analysis and consequences}},}\ }\href@noop {}
  {\bibfield  {journal} {\bibinfo  {journal} {Journal of Chemical Physics}\
  }\textbf {\bibinfo {volume} {129}},\ \bibinfo {pages} {184508} (\bibinfo
  {year} {2008}{\natexlab{b}})}\BibitemShut {NoStop}%
\bibitem [{\citenamefont {{Thomas B. Schr{\o}der and Nicholas P. Bailey and Ulf
  R. Pedersen and Nicoletta Gnan and Jeppe C. Dyre}}(2009)}]{paper3}%
  \BibitemOpen
  \bibfield  {author} {\bibinfo {author} {\bibnamefont {{Thomas B. Schr{\o}der
  and Nicholas P. Bailey and Ulf R. Pedersen and Nicoletta Gnan and Jeppe C.
  Dyre}}},\ }\bibfield  {title} {\enquote {\bibinfo {title} {{Pressure-energy
  correlations in liquids. III. Statistical mechanics and thermodynamics of
  liquids with hidden scale invariance}},}\ }\href@noop {} {\bibfield
  {journal} {\bibinfo  {journal} {Journal of Chemical Physics}\ }\textbf
  {\bibinfo {volume} {131}},\ \bibinfo {pages} {234503} (\bibinfo {year}
  {2009})}\BibitemShut {NoStop}%
\bibitem [{\citenamefont {{Nicoletta Gnan and Thomas B. Schr{\o}der and Ulf R.
  Pedersen and Nicholas P. Bailey and Jeppe C. Dyre}}(2009)}]{paper4}%
  \BibitemOpen
  \bibfield  {author} {\bibinfo {author} {\bibnamefont {{Nicoletta Gnan and
  Thomas B. Schr{\o}der and Ulf R. Pedersen and Nicholas P. Bailey and Jeppe C.
  Dyre}}},\ }\bibfield  {title} {\enquote {\bibinfo {title} {{Pressure-energy
  correlations in liquids. IV. 'Isomorphs' in liquid state diagrams}},}\
  }\href@noop {} {\bibfield  {journal} {\bibinfo  {journal} {Journal of
  Chemical Physics}\ }\textbf {\bibinfo {volume} {131}},\ \bibinfo {pages}
  {234504} (\bibinfo {year} {2009})}\BibitemShut {NoStop}%
\bibitem [{\citenamefont {{Thomas B. Schr{\o}der and Nicoletta Gnan and Ulf R.
  Pedersen and Nicholas Bailey and Jeppe C. Dyre}}(2011)}]{paper5}%
  \BibitemOpen
  \bibfield  {author} {\bibinfo {author} {\bibnamefont {{Thomas B. Schr{\o}der
  and Nicoletta Gnan and Ulf R. Pedersen and Nicholas Bailey and Jeppe C.
  Dyre}}},\ }\bibfield  {title} {\enquote {\bibinfo {title} {{Pressure-energy
  correlations in liquids. V. Isomorphs in generalized Lennard-Jones
  systems}},}\ }\href@noop {} {\bibfield  {journal} {\bibinfo  {journal}
  {Journal of Chemical Physics}\ }\textbf {\bibinfo {volume} {134}},\ \bibinfo
  {pages} {164505} (\bibinfo {year} {2011})}\BibitemShut {NoStop}%
\bibitem [{\citenamefont {{Trond S. Ingebrigtsen and Thomas B. Schr{\o}der and
  Jeppe C. Dyre}}(2012)}]{simpleliquid}%
  \BibitemOpen
  \bibfield  {author} {\bibinfo {author} {\bibnamefont {{Trond S. Ingebrigtsen
  and Thomas B. Schr{\o}der and Jeppe C. Dyre}}},\ }\bibfield  {title}
  {\enquote {\bibinfo {title} {{What is a simple liquid?}}}\ }\href@noop {}
  {\bibfield  {journal} {\bibinfo  {journal} {Physical Review X}\ }\textbf
  {\bibinfo {volume} {2}},\ \bibinfo {pages} {011011} (\bibinfo {year}
  {2012})}\BibitemShut {NoStop}%
\bibitem [{\citenamefont {{Jeppe C. Dyre}}(2014)}]{hiddenscale}%
  \BibitemOpen
  \bibfield  {author} {\bibinfo {author} {\bibnamefont {{Jeppe C. Dyre}}},\
  }\bibfield  {title} {\enquote {\bibinfo {title} {{Hidden Scale Invariance in
  Condensed Matter}},}\ }\href@noop {} {\bibfield  {journal} {\bibinfo
  {journal} {{The Journal of Physical Chemistry B}}\ }\textbf {\bibinfo
  {volume} {118}},\ \bibinfo {pages} {10007} (\bibinfo {year}
  {2014})}\BibitemShut {NoStop}%
\bibitem [{\citenamefont {{Jeppe C. Dyre}}(2016)}]{JeppeBigReview}%
  \BibitemOpen
  \bibfield  {author} {\bibinfo {author} {\bibnamefont {{Jeppe C. Dyre}}},\
  }\bibfield  {title} {\enquote {\bibinfo {title} {{Simple liquids'
  quasiuniversality and the hard-sphere paradigm}},}\ }\href@noop {} {\bibfield
   {journal} {\bibinfo  {journal} {{J. Phys.: Condens. Matter}}\ }\textbf
  {\bibinfo {volume} {28}},\ \bibinfo {pages} {323001} (\bibinfo {year}
  {2016})}\BibitemShut {NoStop}%
\bibitem [{\citenamefont {Malins}\ \emph {et~al.}(2013)\citenamefont {Malins},
  \citenamefont {Eggers},\ and\ \citenamefont {Royall}}]{mal13}%
  \BibitemOpen
  \bibfield  {author} {\bibinfo {author} {\bibfnamefont {A.}~\bibnamefont
  {Malins}}, \bibinfo {author} {\bibfnamefont {J.}~\bibnamefont {Eggers}}, \
  and\ \bibinfo {author} {\bibfnamefont {C.~P.}\ \bibnamefont {Royall}},\
  }\bibfield  {title} {\enquote {\bibinfo {title} {{Investigating Isomorphs
  with the Topological Cluster Classification}},}\ }\href@noop {} {\bibfield
  {journal} {\bibinfo  {journal} {J. Chem. Phys.}\ }\textbf {\bibinfo {volume}
  {139}},\ \bibinfo {pages} {234505} (\bibinfo {year} {2013})}\BibitemShut
  {NoStop}%
\bibitem [{\citenamefont {Abramson}(2014)}]{abr14}%
  \BibitemOpen
  \bibfield  {author} {\bibinfo {author} {\bibfnamefont {E.~H.}\ \bibnamefont
  {Abramson}},\ }\bibfield  {title} {\enquote {\bibinfo {title} {Viscosity of
  fluid nitrogen to pressures of 10 {GPa}},}\ }\href@noop {} {\bibfield
  {journal} {\bibinfo  {journal} {J. Phys. Chem. B}\ }\textbf {\bibinfo
  {volume} {118}},\ \bibinfo {pages} {11792--11796} (\bibinfo {year}
  {2014})}\BibitemShut {NoStop}%
\bibitem [{\citenamefont {Fernandez}\ and\ \citenamefont
  {Lopez}(2014)}]{fer14}%
  \BibitemOpen
  \bibfield  {author} {\bibinfo {author} {\bibfnamefont {J.}~\bibnamefont
  {Fernandez}}\ and\ \bibinfo {author} {\bibfnamefont {E.~R.}\ \bibnamefont
  {Lopez}},\ }\enquote {\bibinfo {title} {{{\rm in} Experimental
  Thermodynamics: Advances in Transport Properties of Fluids}},}\ \ (\bibinfo
  {publisher} {Royal Society of Chemistry},\ \bibinfo {year} {2014})\ Chap.\
  \bibinfo {chapter} {9.3}, pp.\ \bibinfo {pages} {307--317}\BibitemShut
  {NoStop}%
\bibitem [{\citenamefont {Flenner}\ \emph {et~al.}(2014)\citenamefont
  {Flenner}, \citenamefont {Staley},\ and\ \citenamefont {Szamel}}]{fle14}%
  \BibitemOpen
  \bibfield  {author} {\bibinfo {author} {\bibfnamefont {E.}~\bibnamefont
  {Flenner}}, \bibinfo {author} {\bibfnamefont {H.}~\bibnamefont {Staley}}, \
  and\ \bibinfo {author} {\bibfnamefont {G.}~\bibnamefont {Szamel}},\
  }\bibfield  {title} {\enquote {\bibinfo {title} {{Universal Features of
  Dynamic Heterogeneity in Supercooled Liquids}},}\ }\href@noop {} {\bibfield
  {journal} {\bibinfo  {journal} {Phys. Rev. Lett.}\ }\textbf {\bibinfo
  {volume} {112}},\ \bibinfo {pages} {097801} (\bibinfo {year}
  {2014})}\BibitemShut {NoStop}%
\bibitem [{\citenamefont {Prasad}\ and\ \citenamefont
  {Chakravarty}(2014)}]{pra14}%
  \BibitemOpen
  \bibfield  {author} {\bibinfo {author} {\bibfnamefont {S.}~\bibnamefont
  {Prasad}}\ and\ \bibinfo {author} {\bibfnamefont {C.}~\bibnamefont
  {Chakravarty}},\ }\bibfield  {title} {\enquote {\bibinfo {title} {{Onset of
  Simple Liquid Behaviour in Modified Water Models}},}\ }\href@noop {}
  {\bibfield  {journal} {\bibinfo  {journal} {J. Chem. Phys.}\ }\textbf
  {\bibinfo {volume} {140}},\ \bibinfo {eid} {164501} (\bibinfo {year}
  {2014})}\BibitemShut {NoStop}%
\bibitem [{\citenamefont {{Schr{\o}der, Thomas B. and Dyre, Jeppe
  C.}}(2014)}]{Isomorph2.0}%
  \BibitemOpen
  \bibfield  {author} {\bibinfo {author} {\bibnamefont {{Schr{\o}der, Thomas B.
  and Dyre, Jeppe C.}}},\ }\bibfield  {title} {\enquote {\bibinfo {title}
  {{Simplicity of condensed matter at its core: Generic definition of a
  Roskilde-simple system}},}\ }\href {\doibase 10.1063/1.4901215} {\bibfield
  {journal} {\bibinfo  {journal} {Journal of Chemical Physics}\ }\textbf
  {\bibinfo {volume} {141}},\ \bibinfo {eid} {204502} (\bibinfo {year}
  {2014})}\BibitemShut {NoStop}%
\bibitem [{\citenamefont {Buchenau}(2015)}]{buc15}%
  \BibitemOpen
  \bibfield  {author} {\bibinfo {author} {\bibfnamefont {U.}~\bibnamefont
  {Buchenau}},\ }\bibfield  {title} {\enquote {\bibinfo {title} {Thermodynamics
  and dynamics of the inherent states at the glass transition},}\ }\href@noop
  {} {\bibfield  {journal} {\bibinfo  {journal} {J. Non-Cryst. Solids}\
  }\textbf {\bibinfo {volume} {407}},\ \bibinfo {pages} {179--183} (\bibinfo
  {year} {2015})}\BibitemShut {NoStop}%
\bibitem [{\citenamefont {Harris}\ and\ \citenamefont
  {Kanakubo}(2015)}]{har15}%
  \BibitemOpen
  \bibfield  {author} {\bibinfo {author} {\bibfnamefont {K.~R.}\ \bibnamefont
  {Harris}}\ and\ \bibinfo {author} {\bibfnamefont {M.}~\bibnamefont
  {Kanakubo}},\ }\bibfield  {title} {\enquote {\bibinfo {title}
  {Self-diffusion, velocity cross-correlation, distinct diffusion and
  resistance coefficients of the ionic liquid {[BMIM][Tf2N]} at high
  pressure},}\ }\href {\doibase {10.1039/c5cp04277a}} {\bibfield  {journal}
  {\bibinfo  {journal} {Phys. Chem. Chem. Phys.}\ }\textbf {\bibinfo {volume}
  {17}},\ \bibinfo {pages} {23977--23993} (\bibinfo {year} {2015})}\BibitemShut
  {NoStop}%
\bibitem [{\citenamefont {Heyes}\ \emph {et~al.}(2015)\citenamefont {Heyes},
  \citenamefont {Dini},\ and\ \citenamefont {Branka}}]{hey15}%
  \BibitemOpen
  \bibfield  {author} {\bibinfo {author} {\bibfnamefont {D.~M.}\ \bibnamefont
  {Heyes}}, \bibinfo {author} {\bibfnamefont {D.}~\bibnamefont {Dini}}, \ and\
  \bibinfo {author} {\bibfnamefont {A.~C.}\ \bibnamefont {Branka}},\ }\bibfield
   {title} {\enquote {\bibinfo {title} {Scaling of {Lennard-Jones} liquid
  elastic moduli, viscoelasticity and other properties along fluid-solid
  coexistence},}\ }\href {\doibase 10.1002/pssb.201451695} {\bibfield
  {journal} {\bibinfo  {journal} {Phys. Status Solidi (b)}\ }\textbf {\bibinfo
  {volume} {252}},\ \bibinfo {pages} {1514--1525} (\bibinfo {year}
  {2015})}\BibitemShut {NoStop}%
\bibitem [{\citenamefont {Schirmacher}\ \emph {et~al.}(2015)\citenamefont
  {Schirmacher}, \citenamefont {Ruocco},\ and\ \citenamefont
  {Mazzone}}]{sch15}%
  \BibitemOpen
  \bibfield  {author} {\bibinfo {author} {\bibfnamefont {Walter}\ \bibnamefont
  {Schirmacher}}, \bibinfo {author} {\bibfnamefont {Giancarlo}\ \bibnamefont
  {Ruocco}}, \ and\ \bibinfo {author} {\bibfnamefont {Valerio}\ \bibnamefont
  {Mazzone}},\ }\bibfield  {title} {\enquote {\bibinfo {title} {Heterogeneous
  viscoelasticity: A combined theory of dynamic and elastic heterogeneity},}\
  }\href {\doibase 10.1103/PhysRevLett.115.015901} {\bibfield  {journal}
  {\bibinfo  {journal} {Phys. Rev. Lett.}\ }\textbf {\bibinfo {volume} {115}},\
  \bibinfo {pages} {015901} (\bibinfo {year} {2015})}\BibitemShut {NoStop}%
\bibitem [{\citenamefont {Roland}\ \emph {et~al.}(2005)\citenamefont {Roland},
  \citenamefont {Hensel-Bielowka}, \citenamefont {Paluch},\ and\ \citenamefont
  {Casalini}}]{rol05}%
  \BibitemOpen
  \bibfield  {author} {\bibinfo {author} {\bibfnamefont {C.~M.}\ \bibnamefont
  {Roland}}, \bibinfo {author} {\bibfnamefont {S.}~\bibnamefont
  {Hensel-Bielowka}}, \bibinfo {author} {\bibfnamefont {M.}~\bibnamefont
  {Paluch}}, \ and\ \bibinfo {author} {\bibfnamefont {R.}~\bibnamefont
  {Casalini}},\ }\bibfield  {title} {\enquote {\bibinfo {title} {{Supercooled
  Dynamics of Glass-Forming Liquids and Polymers under Hydrostatic
  Pressure}},}\ }\href@noop {} {\bibfield  {journal} {\bibinfo  {journal} {Rep.
  Prog. Phys.}\ }\textbf {\bibinfo {volume} {68}},\ \bibinfo {pages}
  {1405--1478} (\bibinfo {year} {2005})}\BibitemShut {NoStop}%
\bibitem [{\citenamefont {Hansen}\ and\ \citenamefont
  {McDonald}(2013)}]{han13}%
  \BibitemOpen
  \bibfield  {author} {\bibinfo {author} {\bibfnamefont {J.-P.}\ \bibnamefont
  {Hansen}}\ and\ \bibinfo {author} {\bibfnamefont {I.~R.}\ \bibnamefont
  {McDonald}},\ }\href@noop {} {\emph {\bibinfo {title} {{Theory of Simple
  Liquids: With Applications to Soft Matter}}}},\ \bibinfo {edition} {4th}\
  ed.\ (\bibinfo  {publisher} {Academic, New York},\ \bibinfo {year}
  {2013})\BibitemShut {NoStop}%
\bibitem [{\citenamefont {{Yaakov Rosenfeld}}(1999)}]{Rosenfeld1999}%
  \BibitemOpen
  \bibfield  {author} {\bibinfo {author} {\bibnamefont {{Yaakov Rosenfeld}}},\
  }\bibfield  {title} {\enquote {\bibinfo {title} {{A quasi-universal scaling
  law for atomic transport in simple fluids}},}\ }\href
  {http://stacks.iop.org/0953-8984/11/i=28/a=303} {\bibfield  {journal}
  {\bibinfo  {journal} {Journal of Physics: Condensed Matter}\ }\textbf
  {\bibinfo {volume} {11}},\ \bibinfo {pages} {5415} (\bibinfo {year}
  {1999})}\BibitemShut {NoStop}%
\bibitem [{\citenamefont {{L.D. Landau and E.M. Lifshitz}}(1958)}]{LandauLif}%
  \BibitemOpen
  \bibfield  {author} {\bibinfo {author} {\bibnamefont {{L.D. Landau and E.M.
  Lifshitz}}},\ }\href@noop {} {\emph {\bibinfo {title} {{Statistical
  Physics}}}},\ edited by\ \bibinfo {editor} {\bibnamefont {Oxford}}\ (\bibinfo
   {publisher} {Pergamon},\ \bibinfo {year} {1958})\BibitemShut {NoStop}%
\bibitem [{\citenamefont {{L. B{\o}hling and T. S. Ingebrigtsen and A.
  Grzybowski and M. Paluch and J. C. Dyre and T. B.
  Schr{\o}der}}(2012)}]{Lasse2012}%
  \BibitemOpen
  \bibfield  {author} {\bibinfo {author} {\bibnamefont {{L. B{\o}hling and T.
  S. Ingebrigtsen and A. Grzybowski and M. Paluch and J. C. Dyre and T. B.
  Schr{\o}der}}},\ }\bibfield  {title} {\enquote {\bibinfo {title} {{Scaling of
  viscous dynamics in simple liquids: theory, simulation and experiment}},}\
  }\href {\doibase 10.1088/1367-2630/14/11/113035} {\bibfield  {journal}
  {\bibinfo  {journal} {New J. Phys.}\ }\textbf {\bibinfo {volume} {14}},\
  \bibinfo {pages} {113035} (\bibinfo {year} {2012})}\BibitemShut {NoStop}%
\bibitem [{\citenamefont {{Trond S. Ingebrigtsen and Lasse B{\o}hling and
  Thomas B. Schr{\o}der and Jeppe C. Dyre}}(2012)}]{hrhoTrond}%
  \BibitemOpen
  \bibfield  {author} {\bibinfo {author} {\bibnamefont {{Trond S. Ingebrigtsen
  and Lasse B{\o}hling and Thomas B. Schr{\o}der and Jeppe C. Dyre}}},\
  }\bibfield  {title} {\enquote {\bibinfo {title} {{Thermodynamics of condensed
  matter with strong pressure-energy correlations}},}\ }\href@noop {}
  {\bibfield  {journal} {\bibinfo  {journal} {Journal of Chemical Physics}\
  }\textbf {\bibinfo {volume} {136}},\ \bibinfo {pages} {061102} (\bibinfo
  {year} {2012})}\BibitemShut {NoStop}%
\bibitem [{\citenamefont {{Lorenzo Costigliola}}(2016)}]{PhDthesis}%
  \BibitemOpen
  \bibfield  {author} {\bibinfo {author} {\bibnamefont {{Lorenzo
  Costigliola}}},\ }\emph {\bibinfo {title} {{Isomorph theory and
  extensions}}},\ \href {http://dirac.ruc.dk/~lorenzoc/PhD_thesis/} {Ph.D.
  thesis},\ \bibinfo  {school} {Roskilde Universitet} (\bibinfo {year}
  {2016})\BibitemShut {NoStop}%
\bibitem [{\citenamefont {Ingebrigtsen}\ \emph {et~al.}(2012)\citenamefont
  {Ingebrigtsen}, \citenamefont {B{\o}hling}, \citenamefont {Schr{\o}der},\
  and\ \citenamefont {Dyre}}]{ing12a}%
  \BibitemOpen
  \bibfield  {author} {\bibinfo {author} {\bibfnamefont {T.~S.}\ \bibnamefont
  {Ingebrigtsen}}, \bibinfo {author} {\bibfnamefont {L.}~\bibnamefont
  {B{\o}hling}}, \bibinfo {author} {\bibfnamefont {T.~B.}\ \bibnamefont
  {Schr{\o}der}}, \ and\ \bibinfo {author} {\bibfnamefont {J.~C.}\ \bibnamefont
  {Dyre}},\ }\bibfield  {title} {\enquote {\bibinfo {title} {{Thermodynamics of
  Condensed Matter with Strong Pressure-Energy Correlations}},}\ }\href@noop {}
  {\bibfield  {journal} {\bibinfo  {journal} {J. Chem. Phys.}\ }\textbf
  {\bibinfo {volume} {136}},\ \bibinfo {pages} {061102} (\bibinfo {year}
  {2012})}\BibitemShut {NoStop}%
\bibitem [{\citenamefont {Gundermann}\ \emph {et~al.}()\citenamefont
  {Gundermann}, \citenamefont {Pedersen}, \citenamefont {Hecksher},
  \citenamefont {Bailey}, \citenamefont {Jakobsen}, \citenamefont
  {Christensen}, \citenamefont {Olsen}, \citenamefont {Schroder}, \citenamefont
  {Fragiadakis}, \citenamefont {Casalini}, \citenamefont {Roland},
  \citenamefont {Dyre},\ and\ \citenamefont {Niss}}]{Gundermann2011}%
  \BibitemOpen
  \bibfield  {author} {\bibinfo {author} {\bibfnamefont {Ditte}\ \bibnamefont
  {Gundermann}}, \bibinfo {author} {\bibfnamefont {Ulf~R.}\ \bibnamefont
  {Pedersen}}, \bibinfo {author} {\bibfnamefont {Tina}\ \bibnamefont
  {Hecksher}}, \bibinfo {author} {\bibfnamefont {Nicholas~P.}\ \bibnamefont
  {Bailey}}, \bibinfo {author} {\bibfnamefont {Bo}~\bibnamefont {Jakobsen}},
  \bibinfo {author} {\bibfnamefont {Tage}\ \bibnamefont {Christensen}},
  \bibinfo {author} {\bibfnamefont {Niels~B.}\ \bibnamefont {Olsen}}, \bibinfo
  {author} {\bibfnamefont {Thomas~B.}\ \bibnamefont {Schroder}}, \bibinfo
  {author} {\bibfnamefont {Daniel}\ \bibnamefont {Fragiadakis}}, \bibinfo
  {author} {\bibfnamefont {Riccardo}\ \bibnamefont {Casalini}}, \bibinfo
  {author} {\bibfnamefont {C.~Michael}\ \bibnamefont {Roland}}, \bibinfo
  {author} {\bibfnamefont {Jeppe~C.}\ \bibnamefont {Dyre}}, \ and\ \bibinfo
  {author} {\bibfnamefont {Kristine}\ \bibnamefont {Niss}},\ }\bibfield
  {title} {\enquote {\bibinfo {title} {{Predicting the density-scaling exponent
  of a glass-forming liquid from Prigogine-Defay ratio measurements}},}\ }\href
  {\doibase 10.1038/nphys2031} {\bibfield  {journal} {\bibinfo  {journal}
  {{Nat. Phys.}}\ }\textbf {\bibinfo {volume} {7}},\ \bibinfo {pages}
  {816}}\BibitemShut {NoStop}%
\bibitem [{\citenamefont {{Lorenzo Costigliola, Thomas B. Schr{\o}der and Jeppe
  C. Dyre}}(2016)}]{Costigliola2016b}%
  \BibitemOpen
  \bibfield  {author} {\bibinfo {author} {\bibnamefont {{Lorenzo Costigliola,
  Thomas B. Schr{\o}der and Jeppe C. Dyre}}},\ }\bibfield  {title} {\enquote
  {\bibinfo {title} {{Communication: Studies of the Lennard-Jones fluid in 2,
  3, and 4 dimensions highlight the need for a liquid-state 1/d expansion}},}\
  }\href@noop {} {\bibfield  {journal} {\bibinfo  {journal} {The Journal of
  Chemical Physics}\ }\textbf {\bibinfo {volume} {144}},\ \bibinfo {eid}
  {231101} (\bibinfo {year} {2016})}\BibitemShut {NoStop}%
\bibitem [{\citenamefont {{Smit, B. and Frenkel, D.}}(1991)}]{Smit1991}%
  \BibitemOpen
  \bibfield  {author} {\bibinfo {author} {\bibnamefont {{Smit, B. and Frenkel,
  D.}}},\ }\bibfield  {title} {\enquote {\bibinfo {title} {{Vapor-liquid
  equilibria of the two-dimensional Lennard-Jones fluid(s)}},}\ }\href
  {\doibase 10.1063/1.460477} {\bibfield  {journal} {\bibinfo  {journal} {The
  Journal of Chemical Physics}\ }\textbf {\bibinfo {volume} {94}},\ \bibinfo
  {pages} {5663} (\bibinfo {year} {1991})}\BibitemShut {NoStop}%
\bibitem [{\citenamefont {{Potoff, Jeffrey J. and Panagiotopoulos, Athanassios
  Z.}}(1998)}]{Potoff1998}%
  \BibitemOpen
  \bibfield  {author} {\bibinfo {author} {\bibnamefont {{Potoff, Jeffrey J. and
  Panagiotopoulos, Athanassios Z.}}},\ }\bibfield  {title} {\enquote {\bibinfo
  {title} {{Critical point and phase behavior of the pure fluid and a
  Lennard-Jones mixture}},}\ }\href {\doibase 10.1063/1.477787} {\bibfield
  {journal} {\bibinfo  {journal} {The Journal of Chemical Physics}\ }\textbf
  {\bibinfo {volume} {109}},\ \bibinfo {pages} {10914} (\bibinfo {year}
  {1998})}\BibitemShut {NoStop}%
\bibitem [{\citenamefont {{Hloucha, M. and Sandler, S.
  I.}}(1999)}]{Hloucha1999}%
  \BibitemOpen
  \bibfield  {author} {\bibinfo {author} {\bibnamefont {{Hloucha, M. and
  Sandler, S. I.}}},\ }\bibfield  {title} {\enquote {\bibinfo {title} {{Phase
  diagram of the four-dimensional Lennard-Jones fluid}},}\ }\href {\doibase
  10.1063/1.480138} {\bibfield  {journal} {\bibinfo  {journal} {The Journal of
  Chemical Physics}\ }\textbf {\bibinfo {volume} {111}},\ \bibinfo {pages}
  {8043--8047} (\bibinfo {year} {1999})}\BibitemShut {NoStop}%
\end{thebibliography}
\end{document}